\documentclass[aps,pra,amssymb, amsmath,nobibnotes,nobibnotes,reprint,superscriptaddress,showpacs,showkeys]{revtex4-1}
\usepackage[colorlinks, linkcolor=blue, anchorcolor=blue, citecolor=blue]{hyperref}
\usepackage{bm, amsfonts, mathrsfs}
\usepackage{verbatim, psfrag, times, CJK}
\usepackage{graphics, graphicx, color, epsfig}
\usepackage{float}
\usepackage{flafter,natbib}
\usepackage{indentfirst}

\begin{document}

\newcommand{\ket}[1]{| #1 \rangle}
\newcommand{\bra}[1]{\langle #1 |}

\title{Two-atom system as a nano-antenna for mode switching and light routing}

\author{Vassilis E. Lembessis}
\email{vlempesis@ksu.edu.sa}
\author{Anwar Al Rsheed}
\author{Omar M. Aldossary}
\altaffiliation[Also at ]{The National Center for Mathematics and Physics, KACST, P.O. Box 6086, Riyadh 11442, Saudi Arabia}
\affiliation{Department of Physics and Astronomy, College of Science, King Saud University, Riyadh 11451, P.O. Box 2455, Saudi Arabia}
\author{Zbigniew Ficek}
\email{zficek@kacst.edu.sa}
\affiliation{The National Center for Mathematics and Physics, KACST, P.O. Box 6086, Riyadh 11442, Saudi Arabia}

\date{\today}

\begin{abstract}
We determine how a system composed of two nonidentical two-level atoms with different resonance frequencies and different damping rates could work as a nano-antenna for controlled mode switching and light routing. We calculate the angular distribution of the emitted field detected in a far-field zone of the system including the direct interatomic interactions and arbitrary linear dimensions of the system. The calculation is carried out in terms of the symmetric and antisymmetric modes of the two atom system. We find that as long as the atoms are identical, the emission cannot be switched between the symmetric and antisymmetric modes. The switching may occur when the atoms are non-identical and the emission can then be routed to different modes by changing the relative ratio of the atomic frequencies, or damping rates or by a proper tuning of the laser frequency to the atomic resonance frequencies. It is shown that in the case of atoms of different resonance frequencies but equal damping rates, the light routing is independent of the frequency of the driving laser field. It depends only on the sign of the detuning between the atomic resonance frequencies. In the case of atoms of different damping rates, the emission can be switched between different modes by changing the laser frequency from the blue to red detuned from the atomic resonance. The effect of the interatomic interactions is also considered and it is found that in the case of unequal resonance frequencies of the atoms, the interactions slightly modify the visibility of the intensity pattern. The case of unequal damping rates of the atoms is affected rather more drastically, the light routing becoming asymmetric under the dipole-dipole interaction with the enhanced intensities of the modes turned towards the atom of smaller damping rate. 
\end{abstract}


\maketitle

\section{Introduction}\label{sec:intro} 

There has recently been a considerable interest in studying directional properties of optical nano-antennas composed of dielectric or metallic particles that could emit light into a desired direction~\cite{ts08,kk10,cv10,lc11,nk12,jl13}. Particularly interesting is a bimetallic nano-antenna, recently invented by Shegai {\it et al.}~\cite{she11} that is capable to work as a directional frequency filter which scatters light of different colours in opposite directions. The nano-antenna consists of two metallic nano-particles (gold and silver) of different plasmon resonances and separated by a small distance. It was demonstrated experimentally that the antenna, when driven by a white light, can direct red and blue components in opposite directions. Such an antenna could have many practical applications, for example in optical sensing and could be used as directional single photon sources, important for metrology, quantum computation and quantum information processing.

Another kind of nano-particles that could be employed to work as an optical nano-antenna are single two-level atoms. Physically, when two or more atoms are located at a small distance, it is possible to achieve directional scattering through interference between different modes of the electromagnetic field to which the atoms radiate. Many authors have studied the interference effects theoretically~\cite{leh,ag74,tr82,lm83,bb84,ft87,fs90,kc95,wt97,my97,rf98} and also experimentally~\cite{eb93,vb96} in systems involving few atoms, or composed of a large number of atoms confined to a small volume~\cite{dic,es68,re71,gh82,hf04}, or configured in a linear chain~\cite{ch03,mf07,pc08,hz12}, or self-organized along a waveguide~\cite{cc13}. It has been predicted and experimentally demonstrated that the angular distribution of the emitted radiation depends strongly on the number of atoms and the geometry of the emitting system with emission maxima (superradiance) occurring in some directions with an enhanced intensity up to as much $N^{2}I_{0}$, where $N$ is the number of atoms in a sample and $I_{0}$ is the single atom radiation intensity~\cite{sh73,gf76}. In particular, for a line of atoms, a high focussing of the emission along the line axis can be achieved. The focusing increases with an increasing number of atoms and also with a decreasing distance between the atoms~\cite{mf07,pc08,sf06,wz11,wo12}.

In this paper, we address the question of controlled emissive mode switching and light routing in a system of two two-level atoms. Motivated by the experimental work of Shegai {\it et al.}~\cite{she11} on directional colour routing, we study related effects, namely, we consider a system of two nonidentical two-level atoms separated by an arbitrary distance $r_{12}$ and investigate how the system when driven by an external laser field could work as an optical nano-antenna for a controlled switching of the emission between different modes and for  routing light into a desired direction. We work in terms of the collective symmetric and antisymmetric modes of the system that have different angular distributions. We show that the emission can be switched between the symmetric and antisymmetric modes only if the atoms are nonidentical with either different resonance frequencies or unequal damping rates. The emission can be routed to a desired mode by varying either the ratio of the atomic frequencies or the ratio of the damping rates. In the former, the routing is independent of the frequency of the driving laser field, but in the latter it depends strongly on the frequency of the laser.    

The paper is organized as follows. In Sec.~\ref{sec2} we describe in details our model and the geometry of the system. In Sec.~\ref{sec3} we discuss in details the method we use to evaluate the intensity of the emitted light. In Sec.~\ref{sec4} we derive a general formula for the angular distribution of the emitted radiation. Section~\ref{sec5} provides a simple qualitative explanation of the origin of mode switching and light routing. The method of calculation of the steady-state values of the density matrix elements is presented in Sec.~\ref{sec6}. In Sec.~\ref{sec7}, we discuss the angular distribution of the emitted radiation for independent atoms omitting the collective damping and the dipole-dipole interaction. We present analytical results for the intensity of the emitted light which clearly demonstrate the effects of mode switching and light routing in the system. In Sec.~\ref{sec8}, we present numerical results for the angular distribution with the collective damping and the dipole-dipole interaction included. Finally, we summarize  our results in Sec.~\ref{sec9}.

\section{The model}\label{sec2}

\begin{figure}[h]
\begin{center}
\begin{tabular}{c}
\includegraphics[height=5.5cm]{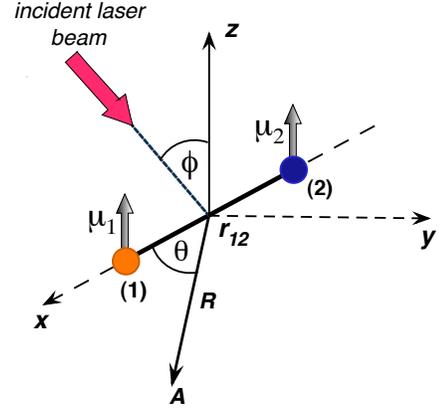}
\end{tabular}
\end{center}
\caption{(Color online) Geometry of the system. Two nonidentical two-level atoms, labeled $(1)$ and $(2)$ are located at fixed positions a distance $r_{12}$ from each other. The atoms are driven by a cw laser field propagating in a direction that is taken to form an angle $\phi$ with the~$z$ axis. The emitted field is detected at point $A$, located on the $xy$ plane distant $R$ from the atoms. The position of the detector on the~$xy$ plane is determined by the angle $\theta$, the direction towards the detecting point relative to the direction of the interatomic~axis.}
\label{fig1} 
\end{figure}

We consider a system composed of two non-identical closely spaced atoms located along the $x$ axis at positions $x_{1}=-\frac{1}{2}r_{12}$ and $x_{2}=\frac{1}{2}r_{12}$, distant $r_{12}$ from each other, as illustrated in Fig.~\ref{fig1}. The atoms are modelled as two-level systems with transitions occurring only between two non-degenerate energy levels $\ket{e_{j}}$ and $\ket{g_{j}}\, (j=1,2)$, having energies $E_{e_{j}}$ and $E_{g_{j}}$ such that $E_{e_{j}} -E_{g_{j}} =\hbar\omega_{j}$, and separated by frequency $\omega_{j}$. We work in the electric dipole approximation that the transitions in the atoms are of the electric dipole type with transition dipole moments $\vec{\mu}_{j}$. It should be pointed out that the model is not restricted to only the electric dipole transitions. It can be extended to any other type of transitions, such as magnetic dipole, electric quadrupole, or two-photon electric dipole transitions. 

The atoms interact with the surrounding (background) three-dimensional electromagnetic field through the electric dipole interaction. The interaction leads to the damping of the atomic transitions with a rate $\Gamma_{j}$ which is equal to the Einstein $A$ coefficient for spontaneous emission. Since the atoms can be at short distances from each other, the interaction of the atoms with the EM field can also lead to a collective behaviour of the atoms that the electromagnetic field produced by an atomic dipole can influence the field produced by the other atomic dipole~\cite{ft02}. In addition, the atoms are continuously driven by a cw monochromatic laser field propagating in the direction both perpendicular to the atomic axis and at an angle $\phi$ to the $z$ axis. The frequency of the laser field $\omega_{L}$ is tuned close to the atomic resonance frequencies with detunings~$\Delta_{1}=\omega_{1}-\omega_{L}$ and $\Delta_{2}=\omega_{2}-\omega_{L}$, respectively. 

The total Hamiltonian of the system composed of two nonidentical atoms driven by a laser field and including the interatomic dipole-dipole interaction can be written as
\begin{eqnarray}
H = H_{0} + H_{L} ,\label{e1}
\end{eqnarray}
where
\begin{eqnarray}
H_{0} =  \hbar \sum_{j=1}^{2}\omega_{j}S_{j}^{z} 
+ \hbar\sum_{i\neq j=1}^{2}\Omega_{ij}S_{i}^{+}S_{j}^{-} \label{e2}
\end{eqnarray}
is the Hamiltonian of the atoms and the dipole-dipole coupling between them, and
\begin{eqnarray}
H_{L} = \frac{1}{2}\hbar\sum_{j=1}^{2}\left[\Omega_{L}(\vec{r}_{j})S_{j}^{+}{\rm e}^{i(\omega_{L}t+\phi_{L})} + {\rm H.c.}\right]  ,\label{e3}
\end{eqnarray}
is the interaction of the atoms with the driven laser field. Here, 
\begin{eqnarray}
\Omega_{L}(\vec{r}_{j}) = \frac{\vec{\mu}_{j}\cdot\vec{E}_{L}}{\hbar}{\rm e}^{i\vec{k}_{L}\cdot \vec{r}_{j}} = \Omega{\rm e}^{i\vec{k}_{L}\cdot \vec{r}_{j}} \label{e4}
\end{eqnarray}
is the position dependent (complex) Rabi frequency of the laser field of the amplitude $\vec{E}_{L}$, initial phase $\phi_{L}$ and the propagation vector~$\vec{k}_{L}$. The operators $S^{+}_{j}=\ket{e_{j}}\bra{g_{j}}$ and $S^{-}_{j}=\ket{g_{j}}\bra{e_{j}}$, appearing in Eq.~(\ref{e2}) are, respectively, the raising and lowering operators for atom $j$, and $S_{j}^{z} = (\ket{e_{j}}\bra{e_{j}} - \ket{g_{j}}\bra{g_{j}})/2$ describes its energy. They are the Pauli spin up and spin down operators for a two-level atom.

The parameter $\Omega_{ij}$ stands for the magnitude of the dipole-dipole interaction between the atoms given by the real part of the interatomic potential, $\Omega_{ij} = {\rm Re}(V_{ij})$, defined as 
\begin{eqnarray}
V_{ij} &=& \sqrt{\Gamma_{i}\Gamma_{j}}\left\{\left(\hat{\mu}_{i}\cdot\hat{\mu}_{j}\right)h_{0}^{(2)}(k_{0}r_{ij})\right. \nonumber\\
&+&\left. \frac{1}{2}\!\left[3\!\left(\hat{\mu}_{i}\cdot\hat{r}_{ij}\right)\!\left(\hat{\mu}_{j}\cdot\hat{r}_{ij}\right) -\hat{\mu}_{i}\cdot\hat{\mu}_{j}\right]\!h_{2}^{(2)}(k_{0}r_{ij})\!\right\} .\label{e5}
\end{eqnarray}
Here $\hat{\mu}_{i}$ and $\hat{r}_{ij}$ are unit vectors in the direction of the $i$th atomic dipole moment $\vec{\mu}_{i}=\mu_{i}\hat{\mu}_{i}$ and in the direction of the atomic axis $\vec{r}_{ij}=r_{ij}\hat{r}_{ij}$, respectively, $k_{0}=\omega_{0}/c$ in which $\omega_{0}=(\omega_{1}+\omega_{2})/2$ is the average frequency of the atomic resonance frequencies, and $h_{n}^{(2)}$ is a spherical Hankel function of the second kind.

\section{Intensity of the emitted field with a two-atom source}\label{sec3}

Our objective is to calculate the intensity $I(\vec{R},t)$ of the field emitted by the system and detected by a single detector at time~$t$ and at an arbitrary point $A$ on the $(x,y)$ plane, $\vec{R}= x\hat{i} +y\hat{j}$, as shown in Fig.~\ref{fig1}. The intensity is proportional to the normally ordered one-time correlation function of the electromagnetic field at the detection point
\begin{eqnarray}
I(\vec{R},t) = u(R)\langle\vec{E}^{(-)}(\vec{R},t)\cdot\vec{E}^{(+)}(\vec{R},t)\rangle ,\label{e6}
\end{eqnarray}
where $\vec{E}^{(+)}(\vec{R},t)$ denotes the positive frequency part of the electromagnetic field detected at the point $\vec{R}$ at time~$t$ and the average is taken over the initial state of the system. The factor $u(R)=(2c\epsilon_{0}R^{2}/\hbar\omega_{0})$ has been introduced so that $I(\vec{R},t)d\Omega_{R}dt$ is the power radiated by the atoms into an element of solid angle $d\Omega_{R}$ around the direction $\vec{R}$ over a small time interval~$dt$ at the moment of time $t$. 

If in addition to the background (free) field there are sources of the EM field such as atoms, the total electric field $\vec{E}(\vec{R},t)$ at the point $A$ can be expressed as the sum of a free-field term~$\vec{E}_{F}(\vec{R},t)$ and the source-field term~$\vec{E}_{S}(\vec{R},t)$:
\begin{eqnarray}
\vec{E}(\vec{R},t) = \vec{E}_{F}(\vec{R},t) + \vec{E}_{S}(\vec{R},t) ,\label{e7}
\end{eqnarray}
where
\begin{eqnarray}
\vec{E}_{F}(\vec{R},t) = i\sum_{k}\!\left(\!\frac{\hbar\omega_{k}}{2\epsilon_{0}V}\!\right)^{\frac{1}{2}}\!\vec{{\rm e}}_{k}a_{k}(0){\rm e}^{i(\vec{k}\cdot\vec{R} -\omega_{k}t)} + {\rm H.c.} ,\label{e8}
\end{eqnarray}
and 
\begin{eqnarray}
\vec{E}_{S}(\vec{R},t) &=& \nabla\times\left[\nabla\times \frac{1}{4\pi\epsilon_{0}}\sum_{j=1}^{2}\frac{\vec{\mu}_{j}(t-|\vec{R}-\vec{r}_{j}|/c)}{|\vec{R}-\vec{r}_{j}|}\right. \nonumber\\
&&\left. \times\theta(t-|\vec{R}-\vec{r}_{j}|/c)\right] .\label{e9}
\end{eqnarray}
The source part of the field is in the retarded form that the field at $(\vec{R},t)$ depends on the dipole moment $\vec{\mu}_{j}$ of the $j$th atom at the retarded time $t-|\vec{R}-\vec{r}_{j}|/c$, where $\vec{r}_{j}$ is the position of the atom, and $\theta$ is the usual Heaviside function, zero for negative argument and unity for positive argument. 

The expression describes the source field for an arbitrary point $(\vec{R},t)$. Usually, we detect fields at large distances from the source atoms, in the so-called far field radiation zone. If the detection point $A$ lies in the far field zone from the atomic system, $R\gg |\vec{r}_{2}-\vec{r}_{1}|$, the source part takes an asymptotic form
\begin{eqnarray}
\vec{E}_{S}(\vec{R},t) &=& \frac{1}{4\pi\epsilon_{0}c^{2}}\sum_{j=1}^{2}\left(\vec{R}-\vec{r}_{j}\right)\nonumber\\
&\times&\left[\frac{\left(\vec{R}-\vec{r}_{j}\right)
\times\ddot{\vec{\mu}}_{j}(t-|\vec{R}-\vec{r}_{j}|/c)}{|\vec{R}-\vec{r}_{j}|^{3}}\right] ,\label{e10}
\end{eqnarray}
where the double dot over $\vec{\mu}_{j}$ stands for the second derivative over time that the source field depends on the dipole acceleration.

The electric dipole operator $\vec{\mu}_{j}$ can be written as the sum of the raising $S^{+}_{j}$ and lowering $S^{-}_{j}$ operators
\begin{eqnarray}
\vec{\mu}_{j}(t) = \vec{p}_{j}S_{j}^{+}(t) +\vec{p}_{j}^{\ast}S_{j}^{-}(t) ,\label{e11}
\end{eqnarray}
where $\vec{p}_{j} = \bra{e_{j}}\vec{\mu_{j}}\ket{g_{j}}$ is the dipole matrix element of the two-level transition in the atom $j$. 

Approximating $S^{\pm}_{j}(t-|\vec{R}-\vec{r}_{j}|/c)$ by their free evolution expressions
\begin{eqnarray}
S^{\pm}_{j}(t-|\vec{R}-\vec{r}_{j}|/c) \approx S^{\pm}_{j}\exp[\pm i(k\hat{R}\cdot\vec{r}_{j} -\omega_{j}t)] ,\label{e12}
\end{eqnarray}
gives $\vec{E}_{S}(\vec{R},t)$ at large distances in terms of the positive and negative frequency components as
\begin{eqnarray}
\vec{E}_{S}(\vec{R},t) = \vec{E}^{(+)}_{S}(\vec{R},t) + \vec{E}^{(-)}_{S}(\vec{R},t) ,\label{e13}
\end{eqnarray}
where
\begin{eqnarray}
E^{(+)}_{S}(\vec{R},t) &=& \frac{-1}{4\pi\epsilon_{0}c^{2}}\sum_{j=1}^{2}
\frac{[\vec{R}\times(\vec{R}\times\vec{p}^{\ast}_{j})]}{R^{3}}\nonumber\\
&\times& \omega_{j}^{2}S_{j}^{-}{\rm e}^{-i(k\hat{R}\cdot \vec{r}_{j}-\omega_{j}t)}  ,\label{e14}
\end{eqnarray}
and
\begin{eqnarray}
E^{(-)}_{S}(\vec{R},t) &=& \frac{-1}{4\pi\epsilon_{0}c^{2}}\sum_{j=1}^{2}
\frac{[\vec{R}\times(\vec{R}\times\vec{p}_{j})]}{R^{3}}\nonumber\\
&\times& \omega_{j}^{2}S_{j}^{+}{\rm e}^{i(k\hat{R}\cdot \vec{r}_{j}-\omega_{j}t)}  .\label{e15}
\end{eqnarray}
In the derivation of the above expressions, we have used the approximation $|\vec{R} -\vec{r}_{j}|\approx \hat{R}\cdot\vec{r}_{j}$, where $\hat{R}=\vec{R}/R$ is the unit vector in the direction of $\vec{R}$. It is seen that the positive (negative) frequency part of the source field $E^{(+)}_{S}(\vec{R},t)\, (E^{(-)}_{S}(\vec{R},t))$ produced by the atoms at the point $\vec{R}$ in far field zone is proportional to the atomic lowering (raising) operators.

In general, the intensity $I(\vec{R},t)$ detected at the point $A$ can be considered in terms of the free-field and the source-field parts by substituting Eq.~(\ref{e13}) into Eq.~(\ref{e6}) which yields
\begin{eqnarray}
&&I(\vec{R},t) = u(R)\left[\langle\vec{E}^{(-)}_{F}(\vec{R},t)\cdot\vec{E}^{(+)}_{F}(\vec{R},t)\rangle\right. \nonumber\\
&&+\left. \langle\vec{E}^{(-)}_{F}(\vec{R},t)\cdot\vec{E}^{(+)}_{S}(\vec{R},t)\rangle + \langle\vec{E}^{(-)}_{S}(\vec{R},t)\cdot\vec{E}^{(+)}_{F}(\vec{R},t)\rangle\right. \nonumber\\
&&+\left. \langle\vec{E}^{(-)}_{S}(\vec{R},t)\cdot\vec{E}^{(+)}_{S}(\vec{R},t)\rangle\right] .\label{e16}
\end{eqnarray}
The intensity equals the sum of the free field and the source field contributions together with interference terms involving both the free field and the source field. In practice, the free field is in the vacuum state $\ket{\{0\}}$, for which
\begin{eqnarray}
\vec{E}^{(+)}_{F}(\vec{R},t)\ket{\{0\}} \equiv 0 ,\label{e17}
\end{eqnarray}
and the detection point $\vec{R}$ is located outside the region of the driving field. We may therefore ignore the contribution of the free-field part and all the interference parts leaving the intensity given by the source part only
\begin{eqnarray}
I(\vec{R},t) = u(R)\langle\vec{E}^{(-)}_{S}(\vec{R},t)\cdot\vec{E}^{(+)}_{S}(\vec{R},t)\rangle .\label{e18}
\end{eqnarray}
Hence, we can express the intensity in terms of the atomic raising and lowering operators by substituting Eq.~(\ref{e14}) and~(\ref{e15}) into Eq.~(\ref{e18}). The intensity is then given by
\begin{eqnarray}
I(\vec{R},t) = u(\vartheta)\sum_{i,j=1}^{2}\sqrt{\Gamma_{i}\Gamma_{j}}\langle S_{i}^{+}(t)S_{j}^{-}(t)\rangle {\rm e}^{ik\hat{R}\cdot\vec{r}_{ij}} ,\label{e19}
\end{eqnarray}
with $u(\vartheta) = (3/8\pi)\sin^{2}\vartheta$, in which $\vartheta$ is the angle between the observation direction $\vec{R}$ and the polarization of the atomic dipole moments $\vec{p}_{i}$. It follows that the correlation functions of the atomic dipole operators $\langle S_{i}^{+}(t)S_{j}^{-}(t)\rangle$ are a measure of the radiation intensity in the far field zone. In the derivation of Eq.~(\ref{e19}) we have assumed that the atomic dipole moments are parallel to each other, $\vec{p}_{1}\parallel \vec{p}_{2}$. This is justified, if one notice that the atomic dipole moments are both induced by the same EM field.

\section{Angular distribution of the emitted field}\label{sec4}

We now turn to perform the summation over $i$ and $j$ in Eq.~(\ref{e19}) and discuss separately the contribution of different terms. This will allow us to extract terms that are responsible for the variation of the intensity with the direction of observation $\vec{R}$. If we perform the summation, we obtain
\begin{eqnarray}
I(\vec{R},t) &=& u(\vartheta)\left\{\Gamma_{1}\langle S_{1}^{+}(t)S_{1}^{-}(t)\rangle +\Gamma_{2}\langle S_{2}^{+}(t)S_{2}^{-}(t)\rangle\right. \nonumber\\
&+&\left. \sqrt{\Gamma_{1}\Gamma_{2}}\left[\langle S_{1}^{+}(t)S_{2}^{-}(t)\rangle\exp(ikr_{12}\cos\theta)\right.\right. \nonumber\\
&+&\left.\left. \langle S_{2}^{+}(t)S_{1}^{-}(t)\rangle\exp(-ikr_{12}\cos\theta)\right]\right\} .\label{e20}
\end{eqnarray}
The physical consequences of the three terms in $I(\vec{R},t)$ are as follows. The first two terms correspond to the radiation emitted by two separate atoms. These two terms are independent of~$\theta$ and therefore they do not vary with the direction of observation $\vec{R}$. The third term, which we shall call the "interference term", is more interesting because it gives rise to variation of the intensity with the direction~$\vec{R}$. It is composed of two terms involving cross correlations between the different atoms, $\langle S_{1}^{+}(t)S_{2}^{-}(t)\rangle$ and $\langle S_{2}^{+}(t)S_{1}^{-}(t)\rangle$. The cross correlations result from the interference between electric fields emitted by the different atoms that the field spontaneously emitted by one of the atoms can be absorbed by the other atom. The interference term can be written as a sum of two terms
\begin{align}
&I_{int}(\vec{R},t) = u(\vartheta)\sqrt{\Gamma_{1}\Gamma_{2}}\left\{\left[\langle S_{1}^{+}(t)S_{2}^{-}(t)\rangle\right.\right. \nonumber\\
&\left.\left. + \langle S_{2}^{+}(t)S_{1}^{-}(t)\rangle\right]\cos(kr_{12}\cos\theta)\right. \nonumber\\
&\left. + i\left(\langle S_{1}^{+}(t)S_{2}^{-}(t)\rangle - \langle S_{2}^{+}(t)S_{1}^{-}(t)\rangle\right)\sin(kr_{12}\cos\theta)\right\} .\label{e21}
\end{align}
Hence, the interference term can be regarded as being made up of the sum of two contributions, one involving a symmetric combination and the other involving an antisymmetric combination of the atomic operators.
There are two kinds of terms that could be interpreted as symmetric and asymmetric modes to which the atoms radiate. As we shall see, these terms may lead to different effects. 

Note that the number of modes and their angular distribution depend on the distance between the atoms. Consider separately the angular distribution of the symmetric and anti-symmetric modes. It is seen from Eq.~(\ref{e21}) that the angular distribution of the symmetric modes is given by a simple relation
\begin{align}
kr_{12}\cos\theta = n\pi, \quad n=0,\pm 1,\pm 2 ,\ldots \label{e22}
\end{align}
or equivalently 
\begin{align}
\cos\theta = \frac{n\lambda}{2r_{12}} ,\quad n=0,\pm 1,\pm 2 ,\ldots ,\label{e23}
\end{align} 
whereas the angular distribution of the anti-symmetric modes is given by
\begin{align}
kr_{12}\cos\theta = \left(n+\frac{1}{2}\right)\pi, \quad n=0,\pm 1,\pm 2,\ldots \label{e24}
\end{align}
which for angles $\theta$ may be written as
\begin{align}
\cos\theta = \frac{\left(n+\frac{1}{2}\right)\lambda}{2r_{12}}  ,\quad n=0,\pm 1,\pm 2,\ldots \label{e25}
\end{align} 
It is evident from Eqs.~(\ref{e23}) and (\ref{e25}) that there is a discrete and a finite number of directions into which the symmetric and antisymmetric modes can make the maximal contribution. Note that the sign of the contributions to the intensity depends on whether $\cos(kr_{12}\cos\theta)$ and $\sin(kr_{12}\cos\theta)$ have positive or negative values. It is apparent by inspection of Eqs.~(\ref{e23}) and~(\ref{e25}) that for even $n$, both terms have positive values, whereas for odd~$n$ they have negative values. Consequently, if the symmetric and antisymmetric combinations of the atomic correlations are positive, maximum values of the intensity will be observed in the directions corresponding to even $n$, and minimum values will be observed in the directions corresponding to odd $n$. 

Consider in some details the directions in which maximum and minimum values of the intensity may be located and whether they correspond to the directions of propagation of symmetric or antisymmetric modes. First, we note that there is no an antisymmetric mode propagating in the direction normal to the atomic axis $(\theta =\pi/2)$. However, there is a symmetric mode propagating in the direction normal to the atomic axis, $\theta=\pi/2$, to which the system radiates for all values of $r_{12}$. For very small distances, $r_{12}<\lambda/4$, the symmetric mode propagating in the direction $\theta=\pi/2$ is the only mode to which the system can radiate. For $r_{12}=\lambda/4$, there are three modes to which the system can radiate, a symmetric mode propagating in the direction $\theta=\pi/2$ and two antisymmetric modes propagating along the atomic axis, one in the direction $\theta=0$ and the other in $\theta=\pi$. As $r_{12}$ increases, the number of modes increases. For example, at $r_{12}=\lambda/2$ the system may radiate into three symmetric modes propagating in directions $\theta=0, \pi/2$ and $\pi$, and into two antisymmetric modes propagating in directions $\theta = \pi/3$ and~$2\pi/3$.  
\begin{figure}[h]
\begin{center}
\begin{tabular}{c}
\includegraphics[height=7.0cm]{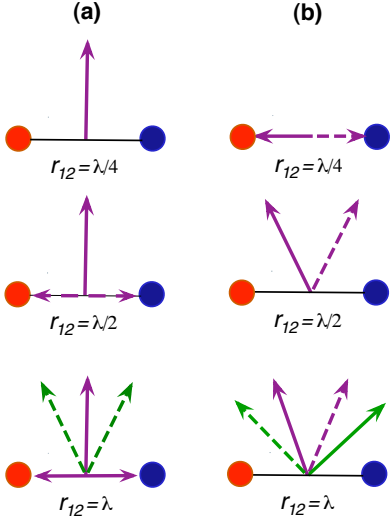}
\end{tabular}
\end{center}
\caption{(Color online) Angular distribution of (a) symmetric and (b) antisymmetric modes for different interatomic separations, $r_{12}=\lambda/4$, $r_{12}=\lambda/2$ and $r_{12}=\lambda$. The solid lines show the directions of the modes to which the interference term contributes positively to produce maxima in the radiation intensity, and the dashed lines show the directions of the modes to which the interference term contributes negatively resulting in minima of the intensity.}
\label{fig2} 
\end{figure}   

Figure~\ref{fig2} shows the angular distribution of the symmetric and antisymmetric modes for three different separations between the atoms, $r_{12}=\lambda/4, \lambda/2$ and $\lambda$. The frame (a) shows the angular distribution of the symmetric modes while the frame (b) shows the distribution of the antisymmetric modes. It follows from the figure that the angular distribution of both symmetric and antisymmetric modes is strictly symmetric about $\theta =\pi/2$. In the case of the symmetric modes, frame (a), and the atomic separation $r_{12}=\lambda/4$, the direction $\theta=\pi/2$ is the only direction the symmetric modes propagate. For $r_{12}=\lambda/2$, the directions $\theta =0$ and $\pi$ for $r_{12}=\lambda/2$ both correspond to a minimum of the intensity. The same property is observed for $r_{12}=\lambda$, where the symmetric modes propagating in the directions $\theta =\pi/3$ and $2\pi/3$ both correspond to a minimum while the modes propagating in the directions $\theta=0$ and $\theta=\pi$ both correspond to a maximum of the intensity. This means that there is no directionality effect if the system radiates through the symmetric modes. This shows, however, that there is an interesting effect of splitting of the emitted radiation into two beams propagating in opposite directions relative to the direction normal to the atomic axis. In both directions, there could be simultaneously a maximum or minimum of the radiation intensity.

The angular distribution of the antisymmetric modes, shown in frame (b), is intrinsically different from those of the symmetric modes. The modes are symmetrically redistributed around the direction normal to the atomic axis, but in each pair of the modes, one of the modes corresponds to a maximum and the other to a minimum of the intensity. This means that depending on the sign of the atomic correlations, the system will radiate either to the left or to the right from the direction normal to the atomic axis. Thus, we see clearly that the antisymmetric modes exhibit the directionality effect. Needless to say, preparing the system to radiate through the antisymmetric modes is the way to achieve a directional emission of the radiation field. 

From this simple analysis it follows that the directionality or light routing can occur in the radiation emitted from the system and, in particular, that these effects are connected to the properties of the antisymmetric modes. It also shows that light routing is strongly pronounced for small interatomic separations and the visibility of this effect decreases with an increasing separation.

\section{Origin of mode switching and light routing}\label{sec5}

We have seen that the interference term involves a sum of two contributions, the symmetric and anti-symmetric combinations of the atomic operators. We now proceed to identify the origin of switching the emission between different modes and the directional light routing, in particular what is required to achieve mode switching and which of these two contributions is responsible for light routing. Since the contributions involve the symmetric and anti-symmetric combinations of the atomic operators, it is convenient to write the intensity in terms of the raising and lowering operators of collective states of the two-atom system that are eigenstates of the Hamiltonian $H_{0}$. In the absence of the interaction between the atoms, the Hilbert space of the two-atom system can be spanned by four product states
\begin{eqnarray}
\ket {g_{1}}\ket {g_{2}} ,\quad \ket {e_{1}}\ket {g_{2}} ,\quad \ket {g_{1}}\ket{e_{2}} ,
\quad \ket {e_{1}}\ket {e_{2}} ,\label{e26}
\end{eqnarray}
and in the basis of these states the Hamiltonian $H_{0}$ can be written in a matrix form~as
\begin{eqnarray}
H_{0} &=& \hbar \left(
\begin{array}{cccc}
-\omega_{0} & 0 & 0 & 0 \\
0 & \frac{1}{2}\Delta & \Omega_{12} & 0 \\
0 & \Omega_{12} & -\frac{1}{2}\Delta & 0 \\
0 & 0 & 0 & \omega_{0}
\end{array}
\right) , \label{e27}
\end{eqnarray}
where $\omega_{0}=(\omega_{1}+\omega_{2})/2$ and $\Delta = (\omega_{1}-\omega_{2})$. It is seen that the matrix is not diagonal due to the presence of the dipole-dipole interaction, and the diagonalization leads to the following energy eigenstates 
\begin{align}
\ket g &= \ket {g_{1}}\ket {g_{2}} ,\quad \ket e = \ket {e_{1}}\ket {e_{2}} ,\nonumber\\
\ket s &= (\sin\alpha)\ket {g_{1}}\ket {e_{2}} + (\cos\alpha)\ket {e_{1}}\ket {g_{2}} ,\nonumber\\
\ket a &= (\cos\alpha)\ket {g_{1}}\ket {e_{2}} -(\sin\alpha)\ket {e_{1}}\ket {g_{2}} ,\label{e28}
\end{align}
where 
\begin{eqnarray}
\cos^{2}\alpha = \frac{1}{2} +\frac{\Delta}{2\sqrt{4\Omega^{2}_{12}+\Delta^{2}}} .\label{e29}
\end{eqnarray}

The states $\ket s$ and $\ket a$ are the symmetric and antisymmetric superpositions of the atomic states, respectively. In fact, these states are non-maximally entangled states of the two-atom system. In the case of identical atoms $(\Delta=0)$, the states reduce to maximally entangled states that are well known in the literature as the Dicke states~\cite{dic}. 

Using Eq.~(\ref{e28}) we find that the atomic raising operators can be expressed in terms of the superposition states as
\begin{align}
S_{1}^{+} &= S_{sg}\cos\alpha - S_{ag}\sin\alpha +S_{es}\sin\alpha + S_{ea}\cos\alpha ,\nonumber\\
S_{2}^{+} &= S_{sg}\sin\alpha + S_{ag}\cos\alpha +S_{es}\cos\alpha - S_{ea}\sin\alpha ,\label{e30}
\end{align}
where $S_{nm} = \ket{n}\bra m$ are the operators of the transitions between the Dicke states, $m,n =a,e,g,s$, and the atomic lowering operators are obtained by taking Hermit conjugate of~Eq.~(\ref{e30}). 

In terms of these new operators, the correlation functions appearing in the interference terms of the intensity become
\begin{align}
\langle S_{1}^{+}(t)S_{2}^{-}(t)\rangle &+ \langle S_{2}^{+}(t)S_{1}^{-}(t)\rangle \nonumber\\
&= \left(\langle S_{ss}(t)\rangle -\langle S_{aa}(t)\rangle\right)\sin(2\alpha) \nonumber\\
&+\left(\langle S_{as}(t)\rangle +\langle S_{sa}(t)\rangle\right)\cos(2\alpha) , \nonumber\\
\langle S_{1}^{+}(t)S_{2}^{-}(t)\rangle &- \langle S_{2}^{+}(t)S_{1}^{-}(t)\rangle \nonumber\\
&= \langle S_{as}(t)\rangle -\langle S_{sa}(t)\rangle ,\label{e31}
\end{align}
from which we see that crucial for the angular variation of the intensity is the presence of a non-zero dipole moment between the symmetric and antisymmetric states. 

For purposes of the physical interpretation, it is convenient to associate the correlation functions with matrix elements of the density operator $\rho$ of the two-atom system. If we use the collective states as the basis states for a matrix representation of the density operator 
\begin{align}
\rho =\sum_{m,n=a,e,g,s}\rho_{mn}\ket n\bra m ,\label{e32}
\end{align}
we arrive at the expressions
\begin{align}
\langle S_{1}^{+}(t)S_{1}^{-}(t)\rangle &= \rho_{ee}(t) + \rho_{ss}(t)\cos^{2}\alpha \nonumber\\
&+\rho_{aa}(t)\sin^{2}\alpha -{\rm Re}[\rho_{sa}(t)]\sin 2\alpha ,\nonumber\\
\langle S_{2}^{+}(t)S_{2}^{-}(t)\rangle &= \rho_{ee}(t) + \rho_{ss}(t)\sin^{2}\alpha \nonumber\\
&+\rho_{aa}(t)\cos^{2}\alpha + {\rm Re}[\rho_{sa}(t)]\sin 2\alpha ,\nonumber\\
\langle S_{1}^{+}(t)S_{2}^{-}(t)\rangle &+ \langle S_{2}^{+}(t)S_{1}^{-}(t)\rangle \nonumber\\ 
&= \left[\rho_{ss}(t) -\rho_{aa}(t)\right]\sin 2\alpha \nonumber\\
&+2{\rm Re}[\rho_{sa}(t)]\cos 2\alpha ,\nonumber\\
\langle S_{1}^{+}(t)S_{2}^{-}(t)\rangle &- \langle S_{2}^{+}(t)S_{1}^{-}(t)\rangle = 2{\rm Im}[\rho_{sa}(t)] ,\label{e33}
\end{align}
where $\rho_{ss}(t)$ and $\rho_{aa}(t)$ are, respectively, the populations of the symmetric and antisymmetric states, and $\rho_{sa}(t)$ is the coherence between these states. 

We may therefore write the radiation intensity as
\begin{align}
I(\vec{R},t) &= u(\vartheta)\Gamma_{0}\left\{I_{0}(t) + I_{c}(t)\cos(kr_{12}\cos\theta)\right. \nonumber\\
&\left. + I_{s}(t)\sin(kr_{12}\cos\theta)\right\} ,\label{e34}
\end{align}
in which $\Gamma_{0} =(\Gamma_{1}+\Gamma_{2})/2$ is the average damping rate of the atoms, $\gamma = (\Gamma_{1}-\Gamma_{2})/(\Gamma_{1}+\Gamma_{2})$ stands for a normalized difference between the damping rates, and we have divided the radiation intensity into three separate parts. The first part~$I_{0}(t)$, which can be expressed in the form
\begin{align}
I_{0}(t) &= \rho_{ee}(t) +\frac{1}{2}(1+\gamma\cos2\alpha)\rho_{ss}(t) \nonumber\\
&+\frac{1}{2}(1-\gamma\cos2\alpha)\rho_{aa}(t) -\gamma {\rm Re}[\rho_{sa}(t)]\sin2\alpha \label{e35}
\end{align}
represents a fraction of the radiation intensity which is independent of the direction of observation. It can be regarded as the background radiation intensity. The second part
\begin{align}
I_{c}(t) &= \sqrt{1-\gamma^{2}}\left\{\left[\rho_{ss}(t) -\rho_{aa}(t)\right]\sin2\alpha\right. \nonumber\\
&+\left. 2{\rm Re}[\rho_{sa}(t)]\!\cos2\alpha\right\} \label{e36}
\end{align}
represents a fraction of the intensity that varies with the direction of observation as $\cos(kr_{12}\cos\theta)$, and
\begin{align}
I_{s}(t) = \sqrt{1-\gamma^{2}}\, {\rm Im}[\rho_{sa}(t)] \label{e37}
\end{align}
is a fraction of the intensity that varies with the direction of observation as $\sin(kr_{12}\cos\theta)$. 

Evidently, every part that contributes to the radiation intensity can be analysed separately and is known once the density matrix elements are determined. Instead of working in terms of the two interference terms one can combine them into a single term, an effective interference term, and then the radiation intensity can be written as
\begin{align}
I(\vec{R},t) = u(\vartheta)\Gamma_{0}\!\left[I_{0}(t) + I_{e}(t)\cos(\psi - kr_{12}\cos\theta)\right] ,\label{e37a}
\end{align}
where 
\begin{align}
I_{e}(t) = \frac{I_{c}(t)}{\cos\psi} \quad {\rm and}\quad \tan\psi =\frac{I_{s}(t)}{I_{c}(t)} .\label{e37b}
\end{align}
As can be seen from Eq.~(\ref{e37a}), the sine term introduces a phase shift of the interference pattern. For $\psi=0$ directions of the maxima and minima are determined by $\cos(kr_{12}\cos\theta)$ and change to that determined by $\sin(kr_{12}\cos\theta)$ when $\psi$ varies from zero to $\pi/2$. Thus, switching between symmetric and the antisymmetric modes can be interpreted as resulting from the phase shift from $\psi=0$ to $\psi=\pi/2$.   

The phase shift is determined by the amplitudes $I_{c}(t)$ and $I_{s}(t)$. 
It is clear from the form of~$I_{c}(t)$, given by Eq.~(\ref{e36}), that the cosine term makes a nonzero contribution to the intensity only when the states~$\ket s$ and~$\ket a$ are unequally populated, $\rho_{ss}(t) -\rho_{aa}(t)\neq 0$, and/or there is a coherence between the states with a nonzero real part,~${\rm Re}[\rho_{sa}(t)]$. Mode switching and light routing is determined by $I_{s}(t)$, given by Eq.~(\ref{e37}), that it is possible only when the coherence $\rho_{sa}(t)$ has a non-vanishing imaginary part, ${\rm Im}[\rho_{sa}(t)]\neq 0$. Note that the real part of the coherence contributes to $I_{c}(t)$ only if $\Delta\neq 0$, as $\cos2\alpha\rightarrow 0$ when $\Delta\rightarrow 0$. On the other hand, $I_{c}(t)$ depends solely on the real part of the coherence when the atoms are independent of each other, as $\sin2\alpha\rightarrow 0$ when $\Omega_{12}\rightarrow 0$. 
\begin{figure}[ht]
\begin{center}
\begin{tabular}{c}
\includegraphics[height=4.5cm]{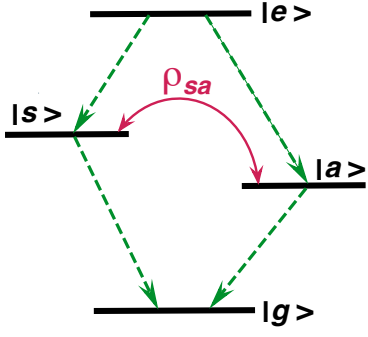}
\end{tabular}
\end{center}
\caption{(Color online) Collective energy states of the system. The dashed lines indicate the possible spontaneous transitions between the states. Essential for mode switching and directional emission is the presence of the coherence~$\rho_{sa}$ between the intermediate states $\ket s$ and $\ket a$.}
\label{fig3} 
\end{figure} 

On the basis of these observations, one can conclude that the presence of the coherence $\rho_{sa}(t)$ is essential to account for mode switching between the symmetric and antisymmetric modes and light routing along the antisymmetric modes. The coherence $\rho_{sa}$ necessary for the mode switching is indicated in~Fig.~\ref{fig3} that shows the collective energy levels of the two-atom system with the possible spontaneous transitions between them.

\section{Evaluation of the density matrix elements}\label{sec6}

The density matrix elements which are needed to evaluate the radiation intensity $I(\vec{R},t)$ are readily calculated from the master equation of the density operator of the system, which has the form~\cite{leh,ag74,ft02}
\begin{align}
\frac{\partial\rho}{\partial t} = &-\frac{i}{\hbar}\left[H,\rho\right] -\frac{1}{2}\Gamma_{1}\left(\rho S_{1}^{+}S_{1}^{-} +S_{1}^{+}S_{1}^{-}\rho -2S_{1}^{-}\rho S_{1}^{+}\right) \nonumber\\
& - \frac{1}{2}\Gamma_{2}\left(\rho S_{2}^{+}S_{2}^{-} +S_{2}^{+}S_{2}^{-}\rho -2S_{2}^{-}\rho S_{2}^{+}\right) \nonumber\\
& - \frac{1}{2}\Gamma_{12}\left(\rho S_{1}^{+}S_{2}^{-} +S_{1}^{+}S_{2}^{-}\rho -2S_{2}^{-}\rho S_{1}^{+}\right) \nonumber\\
& - \frac{1}{2}\Gamma_{12}\left(\rho S_{2}^{+}S_{1}^{-} +S_{2}^{+}S_{1}^{-}\rho -2S_{1}^{-}\rho S_{2}^{+}\right) ,\label{e38}
\end{align}
where $\Gamma_{1}$ and $\Gamma_{2}$ are the spontaneous emission rates of the atom $1$ and $2$, respectively, and $\Gamma_{12}$ is the collective damping rate arising from the mutual coupling of the atoms through the vacuum field. The magnitude of $\Gamma_{12}$ is given by the imaginary part of the interatomic potential, i.e. $\Gamma_{12}={\rm Im}(V_{12})$.

The master equation can be solved and the evolution of the system completely determined by projection of the density operator of the system onto any complete set of basis states. In general, it is a complicated problem since in the basis of the collective states (\ref{e26}), the master equation leads to a set of fifteen coupled differential equations for the density matrix elements that have to be evaluated, in principle, an $15\times 15$ matrix to be diagonalized. It involves twelve off-diagonal and three diagonal matrix elements. The remaining diagonal density matrix element is found from the trace property of the density matrix,~Tr$(\rho) =1$.

Thus, using the master equation (\ref{e38}) and the collective basis (\ref{e26}), we obtain a closed set of fifteen coupled differential equations describing the evolution of the density matrix elements. In a matrix notation, the system of equations can be written as the inhomogeneous equation 
\begin{equation}
\frac{d}{dt}{{\bf Y}}  =\mbox{M}{\bf Y}  +{\bf P} ,\label{e39}
\end{equation}
where M is the $15\times 15$ matrix of the coefficients of the differential equations for the density matrix elements, ${{\bf Y}}$ is a column vector with the following components
\begin{eqnarray}
{{\bf Y}} &=& {\rm col}\left[\rho_{aa},\rho_{ae},\rho_{ag},\rho_{as},\rho_{ea},\rho_{ee},\right. \nonumber\\
&&\left. \rho_{eg},\rho_{es},\rho_{ga},\rho_{ge},\rho_{gg},\rho_{gs},\rho_{sa},\rho_{se},\rho_{sg}\right] ,\label{e40}
\end{eqnarray}
and the column vector ${\bf P}$ has nonzero components
\begin{eqnarray}
P_{4} &=& P_{13} = -\frac{1}{2}i\Delta ,\quad P_{8} = P_{12} = 2i\Omega_{\beta} ,\nonumber\\
P_{11} &=& 2(\sqrt{\Gamma_{1}\Gamma_{2}} +\Gamma_{12}) ,\quad P_{14}=P_{15}= -2i\Omega_{\beta} ,\label{e41}
\end{eqnarray}
where $\Omega_{\beta} = \frac{1}{2}(\Omega_{L1}+\Omega_{L2})$ is the average Rabi frequency of the laser field driving the atoms. 

The matrix equation (\ref{e39}) is a simple differential equation with time independent coefficients, and is solved by direct integration. For an arbitrary initial time $t_0$, the integration of Eq.~(\ref{e39}) leads to the following formal solution 
\begin{equation}
{\bf Y}\left(  t\right) ={\bf Y}\left( t_{0}\right) {\rm e}^{{\rm M}t} -\left(1-{\rm e}^{{\rm M}t}\right){\rm M}^{-1}{\bf P} .\label{e42}
\end{equation}
Because the determinant of the matrix M is different from zero, there exists an inverse matrix M$^{-1}$, so that 
the steady-state values of the components of the vector ${\bf Y}(t)$ can be found by setting the left-hand side of Eq. (\ref{e39}) equal to zero. Thus, the steady-state solution for the components of $Y(\infty)$ is given~by
\begin{equation}
Y_{i}\left(\infty\right)  =-\sum_{j=1}^{15}\left({\rm M}^{-1}\right)_{ij}P_{j} .\label{e43}
\end{equation}
The solutions (\ref{e43}) are technically difficult to describe analytically, but can be evaluated numerically. The solutions are functions of the atomic parameters, $\Delta$ and $\Gamma_{1}, \Gamma_{2}$, the collective parameters $\Omega_{12}$ and $\Gamma_{12}$, and the driving laser parameters, $\Omega_{L1}, \Omega_{L2}$ and the detunings $\Delta_{1}, \Delta_{2}$.  

Before proceeding to a detailed discussion of the angular distribution of the radiation intensity, let us look into the detailed form of the equation of motion for the coherence $\rho_{sa}$ that is required for mode switching and light routing. The equation of motion can be easily determined from the master equation~(\ref{e38})~as
\begin{align}
\dot{\rho}_{sa} =& -(\Gamma_{0} +iU)\rho_{sa} \nonumber\\
&+\frac{1}{2}\Gamma_{0}\left(\gamma\sin 2\alpha -\gamma_{12}\cos 2\alpha\right)(\rho_{ss} +\rho_{aa}) \nonumber\\
&+ \Gamma_{0}\left(\gamma\sin 2\alpha +\gamma_{12}\cos 2\alpha\right)\rho_{ee} ,\label{e44}
\end{align}
where $U=\sqrt{4\Omega_{12}^{2} +\Delta^{2}}$ and $\gamma_{12}=\Gamma_{12}/\Gamma_{0}$.

We first observe from Eq.~(\ref{e44}) that, because $\sin 2\alpha =2\Omega_{12}/U$ and $\cos 2\alpha =\Delta/U$, the coupling of the coherence to the populations is possible only if the atoms are nonidentical and interact with each other. If either atoms are identical or are independent of each other, no coupling of the coherence to the populations is present. In this case, the coherence is created only by the driving field. However, the driving field should be kept weak in order to minimise population of the upper state $\ket e$. It is easy to understand, as predicted by Eq.~(\ref{e34}), the population $\rho_{ee}$ contributes only to the background part of the intensity, $I_{0}(t)$. A large background could make the contributions of the interference terms $I_{c}(t)$ and $I_{s}(t)$ less visible. 

\section{Mode switching and light routing by independent atoms}\label{sec7}

Let us first discuss the mode switching and light routing effects by independent atoms. We assume that the atoms are at a fixed distance $r_{12}$, so there is a fixed phase relation between the atomic dipole moments, but there is no direct exchange of the excitation between them, i.e., the collective damping $\Gamma_{12}$ and the dipole-dipole interaction $\Omega_{12}$ are both equal to zero, $\Gamma_{12}=0$ and $\Omega_{12}=0$. One can notice that in this case the problem simplifies to that of single two-level atoms driven by a coherent laser field. For independent atoms, we can assume that the averages $\langle S_{1}^{+}(t)S_{2}^{-}(t)\rangle$ and $\langle S_{2}^{+}(t)S_{1}^{-}(t)\rangle$, that appear in the expression for the radiation intensity, Eq.~(\ref{e20}),  could be factorized so that $\langle S_{1}^{+}(t)S_{2}^{-}(t)\rangle = \langle S_{1}^{+}(t)\rangle\langle S_{2}^{-}(t)\rangle$ and $\langle S_{2}^{+}(t)S_{1}^{-}(t)\rangle = \langle S_{2}^{+}(t)\rangle\langle S_{1}^{-}(t)\rangle$. In physical terms, the factorization is equivalent to ignore the effect of quantum fluctuations on the mutual correlations between the atoms, but their oscillations still can be kept synchronized by the definite phase of the incident laser field. This is equivalent to assume that the atoms behave mutually coherent although the radiation emitted by each atom is not coherent, i.e. $\langle S_{1}^{+}(t)S_{1}^{-}(t)\rangle\neq \langle S_{1}^{+}(t)\rangle\langle S_{1}^{-}(t)\rangle$ and $\langle S_{2}^{+}(t)S_{2}^{-}(t)\rangle\neq \langle S_{2}^{+}(t)\rangle\langle S_{2}^{-}(t)\rangle$.

The interaction of a two-level atom with a coherent laser field has been extensively studied in the literature~\cite{km76,cw76}. We shall make use of some of the results in these papers, particularly the solutions for the averages involved in the radiation intensity formula, Eq.~(\ref{e20}). If we examine the radiation intensity in the steady-state, the averages required are of the form~\cite{km76,cw76}
\begin{align}
\lim_{t\rightarrow\infty}\langle S_{1}^{+}(t)S_{1}^{-}(t)\rangle &= \Omega^{2}_{L1}/D_{1} ,\nonumber\\
\lim_{t\rightarrow\infty}\langle S_{2}^{+}(t)S_{2}^{-}(t)\rangle &= \Omega^{2}_{L2}/D_{2} ,\nonumber\\
\lim_{t\rightarrow\infty}\langle S_{1}^{+}(t)\rangle &= -\Omega_{L1}\left(\Gamma_{1} - 2i\Delta_{1}\right)/D_{1} ,\nonumber\\
\lim_{t\rightarrow\infty}\langle S_{2}^{-}(t)\rangle &= -\Omega_{L2}\left(\Gamma_{2} + 2i\Delta_{2}\right)/D_{2} ,\label{e45}
\end{align}
with
\begin{align}
D_{1} = 2\Omega^{2}_{L1}\!+\!\Gamma_{1}^{2}\!+\!4\Delta_{1}^{2} ,\quad
D_{2} = 2\Omega^{2}_{L2}\!+\!\Gamma_{2}^{2}\!+\!4\Delta_{2}^{2} ,\label{e46}
\end{align}
where $\Omega_{Li}\, (i=1,2)$ is the Rabi frequency of the laser field at the position of the $i$th atom, $\Delta_{1}= \omega_{1}-\omega_{L}$ and $\Delta_{2}=\omega_{2}-\omega_{L}$ are detunings of the laser field frequency from resonance frequencies of the atoms~$1$ and $2$, respectively. It is easily verified from Eq.~(\ref{e45}) that $\langle S_{1}^{+}(t)S_{1}^{-}(t)\rangle\neq \langle S_{1}^{+}(t)\rangle\langle S_{1}^{-}(t)\rangle$ and $\langle S_{2}^{+}(t)S_{2}^{-}(t)\rangle\neq \langle S_{2}^{+}(t)\rangle\langle S_{2}^{-}(t)\rangle$. In other words, despite the coherent nature of the driving field the emitted radiation by the independent atoms is not coherent in the steady-state.

Substituting Eqs.~(\ref{e45}) into Eq.~(\ref{e20}), we find that the steady-state radiation intensity is of the form
\begin{align}
&I(\vec{R})\equiv \lim_{t\rightarrow\infty}I(\vec{R},t) = \frac{u(\vartheta)\Gamma_{0}\Omega^{2}}{D_{1}D_{2}}\!\left\{(1+\gamma)D_{2}\!+\!(1-\gamma)D_{1}\right. \nonumber\\
&+\left. 2\sqrt{1-\gamma^{2}}\left[\left(4\Delta_{L}^{2} -\Delta^{2} +\Gamma_{0}^{2}(1-\gamma^{2})\right)\cos\left(kr_{12}\cos\theta\right)\right.\right. \nonumber\\
&+\left.\left. 2\Gamma_{0}\left(2\gamma\Delta_{L} -\Delta\right)\sin\left(kr_{12}\cos\theta\right)\right]\right\} ,\label{e47}
\end{align}
where we have translated the detunings $\Delta_{1}$ and~$\Delta_{2}$ to a frequency scale centered on the average atomic frequency~$\omega_{0}$, i.e. $\Delta_{1}=\Delta_{L}+\Delta/2$ and $\Delta_{2}=\Delta_{L}-\Delta/2$, where $\Delta_{L}=\omega_{0}-\omega_{L}$. The damping rates $\Gamma_{1}$ and $\Gamma_{2}$ have also been defined relative to the average damping rate of the atoms~$\Gamma_{0}$. In writing Eq.~(\ref{e47}) we have assumed that the laser field drives the system along the symmetric mode propagating in the direction normal to the atomic axis (Fig.~\ref{fig2}). In this case, $\vec{k}_{L}\cdot\vec{r}_{12}=0$ and  then $\Omega_{L1}=\Omega_{L2}=\Omega$. The situation of driving the system along one of the antisymmetric modes is obtained from Eq.~(\ref{e47}) simply by interchanging $\cos(kr_{12}\cos\theta)$ with~$\sin(kr_{12}\cos\theta)$.  

The variation of the radiation intensity with the direction of observation is provided by the third term in Eq.~(\ref{e47}), and we now proceed to discuss conditions for mode switching and light routing. As an example three  cases of $r_{12}=\lambda/4, \lambda/2$ and $\lambda$ are investigated in details. It was pointed out in Sec.~\ref{sec5} that the occurrence of mode switching and light routing by a system of two interacting atoms require a nonzero amplitude of the sine term in the radiation intensity, Eq.~(\ref{e34}). The same conclusion applies to the case of independent atoms considered here. 
   \begin{figure}[h]
   \begin{center}
   \begin{tabular}{c}
   \includegraphics[height=5.5cm]{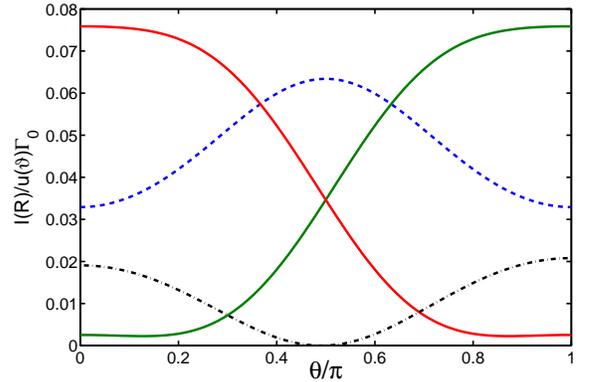}
   \end{tabular}
   \end{center}
   \caption[example]{(Color online) The radiation intensity as a function of the observation direction $\theta$ for the atomic separation $r_{12} =\lambda/4$, $\Omega=0.2\Gamma_{0}$, $\Gamma_{1}=\Gamma_{2}$, $\Delta_{L}=0$ and different values of $\Delta$: $\Delta =0$ (dashed blue line), $\Delta=0.5\Gamma_{0}$ (green solid line), $\Delta=-0.5\Gamma_{0}$ (red solid line), and $\Delta=20\Gamma_{0}$ (dashed-dotted black line, $\times 10^{2}$).}
\label{fig4} 
\end{figure} 

Figure~\ref{fig4} shows the angular distribution of the radiation intensity for the atomic separation $r_{12} =\lambda/4$ and several different detunings $\Delta$. We take $\Gamma_{2}=\Gamma_{1}$ and assume that the atoms are driven by a weak field, $\Omega=0.2\Gamma_{0}$. For identical atoms, $\Delta=0$, and then the intensity exhibits a pronounced peak in the direction $\theta=\pi/2$. As discussed above, $\theta=\pi/2$ is the direction of propagation of the symmetric mode. For $\Delta\neq 0$, we see the switching effect; a transfer of the excitation from the symmetric mode to the antisymmetric modes propagating along the atomic axis, $\theta=0$ and~$\pi$. There is a strong asymmetry in the intensity of the antisymmetric modes with the direction of the enhanced emission dependent on the sign of $\Delta$. For a positive~$\Delta$, the system radiates strongly into the mode propagating in the direction $\theta=\pi$ with almost no emission into the mode $\theta=0$. The direction of the emission reverses when $\Delta\rightarrow -\Delta$. The asymmetry persists for small and moderate $\Delta$ at which, as one can see from Fig.~\ref{fig4}, there still is a nonzero emission into the symmetric mode. When the excitation is completely transferred from the symmetric to the antisymmetric modes, that $I(\vec{R})=0$ at $\theta=\pi/2$, the radiation along $\theta=0$ and $\pi$ becomes symmetric. 

Consider now the dependence of the mode switching and light routing on the frequency of the driving laser. As predicted by Eq.~(\ref{e47}), the amplitude of the cosine term depends on the square of the laser detuning. As such, it does not change the sign when going from blue $(\omega_{L}>\omega_{0})$ to red $(\omega_{L}<\omega_{0})$ detuned cases. That is also consistent with our conclusions in Sec.~\ref{sec5}, where we discussed conditions for light routing of a collective two-atom system. On the other hand, the amplitude of the sine term depends on $\Delta_{L}$ and there is a threshold value for the laser detuning, $\Delta_{L}=\Delta/(2\gamma)$, at which the amplitude reverses sign once we move from $\Delta_{L}<\Delta/(2\gamma)$ to $\Delta_{L}>\Delta/(2\gamma)$. Then the light routing effect is to be expected that for a given direction of observation, the emitted light intensity will be enhanced when $\omega_{L}<\omega_{0}-\Delta/(2\gamma)$, and will be reduced when $\omega_{L}>\omega_{0}-\Delta/(2\gamma)$. In other words, depending on the frequency of the driving field, the emission into the antisymmetric modes can be switched between the~modes. 
   \begin{figure}[h]
   \begin{center}
   \begin{tabular}{c}
   \includegraphics[height=5.8cm]{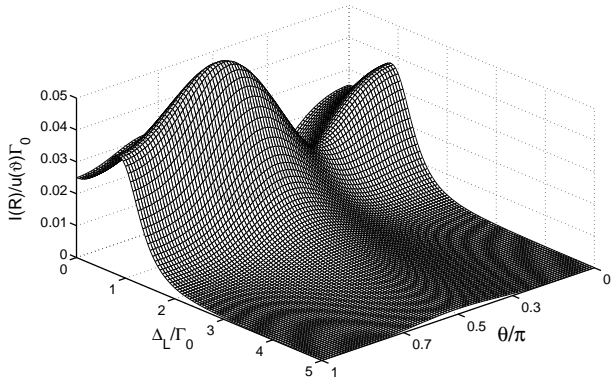}
   \end{tabular}
   \end{center}
   \caption[example]{(Color online) The angular distribution of the steady-state radiation intensity and its variation with the laser detuning $\Delta_{L}$ for $\Omega=0.2\Gamma_{0}$, $r_{12}=\lambda/2$, $\Gamma_{1}=\Gamma_{2}$ and $\Delta=2\Gamma_{0}$.}
\label{fig5} 
\end{figure} 

As we have already mentioned, for $\gamma=0$, i.e. $\Gamma_{1}=\Gamma_{2}$, the switching is independent of $\Delta_{L}$, it depends solely on the sign of $\Delta$. Thus, no light routing dependent on the frequency of the driving laser could be seen in the emitted light if the atoms have the same damping rates, $\Gamma_{1}=\Gamma_{2}$. It follows that $\gamma\neq 0\, (\Gamma_{1}\neq\Gamma_{2})$ is the condition for the frequency dependent light routing. However, the effect of light switching between the symmetric and antisymmetric modes could be observed even if $\gamma=0$ provided that $\Delta\neq 0$. This feature is easily seen in Fig.~\ref{fig5}, which shows the effect of an increasing $\Delta_{L}$ on the angular distribution of the radiation intensity for $r_{12}=\lambda/2$ and $\Delta=2\Gamma_{0}$. For small $\Delta_{L}$ there is a pronounced peak in the direction $\theta =2\pi/3$ and a dip at $\theta=\pi/3$. As $\Delta_{L}$ increases, the intensity distribution turns to a single peak in the direction $\theta=\pi/2$. The angle $\theta=2\pi/3$ corresponds to the direction of propagation of an antisymmetric mode, whereas $\theta=\pi/2$ corresponds to the direction of propagation of a symmetric mode. Thus, for small $\Delta_{L}$, the system emits along the antisymmetric modes. As $\Delta_{L}$ increases, the emission switches from the antisymmetric to the symmetric modes.

The switching of the emission from the antisymmetric to symmetric modes with increasing $\Delta_{L}$ can be understood by examining the analytical formula for the radiation intensity, Eq.~(\ref{e47}). Setting~$\gamma=0$ and $r_{12}=\lambda/2$ in Eq.~(\ref{e47}), it is straightforward to see that the radiation intensity becomes
\begin{align}
I(\vec{R}) &= \frac{2u(\vartheta)\Omega^{2}\Gamma_{0}}{D_{1}D_{2}}\left\{\left(2\Omega^{2}+\Gamma_{0}^{2}+4\Delta_{L}^{2}+\Delta^{2}\right)\right. \nonumber\\
&+\left. \left(4\Delta_{L}^{2} -\Delta^{2} +\Gamma_{0}^{2}\right)\cos\left(\pi\cos\theta\right)\right. \nonumber\\
&\left. - 2\Gamma_{0}\Delta\sin\left(\pi\cos\theta\right)\right\} .\label{e48}
\end{align}
When $\Delta_{L}\leq \Gamma_{0}$, we see that the amplitude of the sine term, representing the contribution of the antisymmetric modes, dominates over the amplitude of the cosine term, representing the contribution of the symmetric modes. Thus, for $\Delta_{L}\leq \Gamma_{0}$ the variation of the intensity with $\theta$ is determined by the sine term. Then at $\theta=\pi/3$, where $\sin(\pi\cos\theta)=1$, the intensity exhibits a dip and a peak at $\theta=2\pi/3$, where $\sin(\pi\cos\theta)=-1$. When $\Delta_{L}>\Gamma_{0}$, the amplitude of the cosine term dominates over that of the sine term and then the intensity exhibits a peak centered at $\theta=\pi/2$, where $\cos(\pi\cos\theta)=1$. 
   \begin{figure}[h]
   \begin{center}
   \begin{tabular}{c}
   \includegraphics[height=5.5cm]{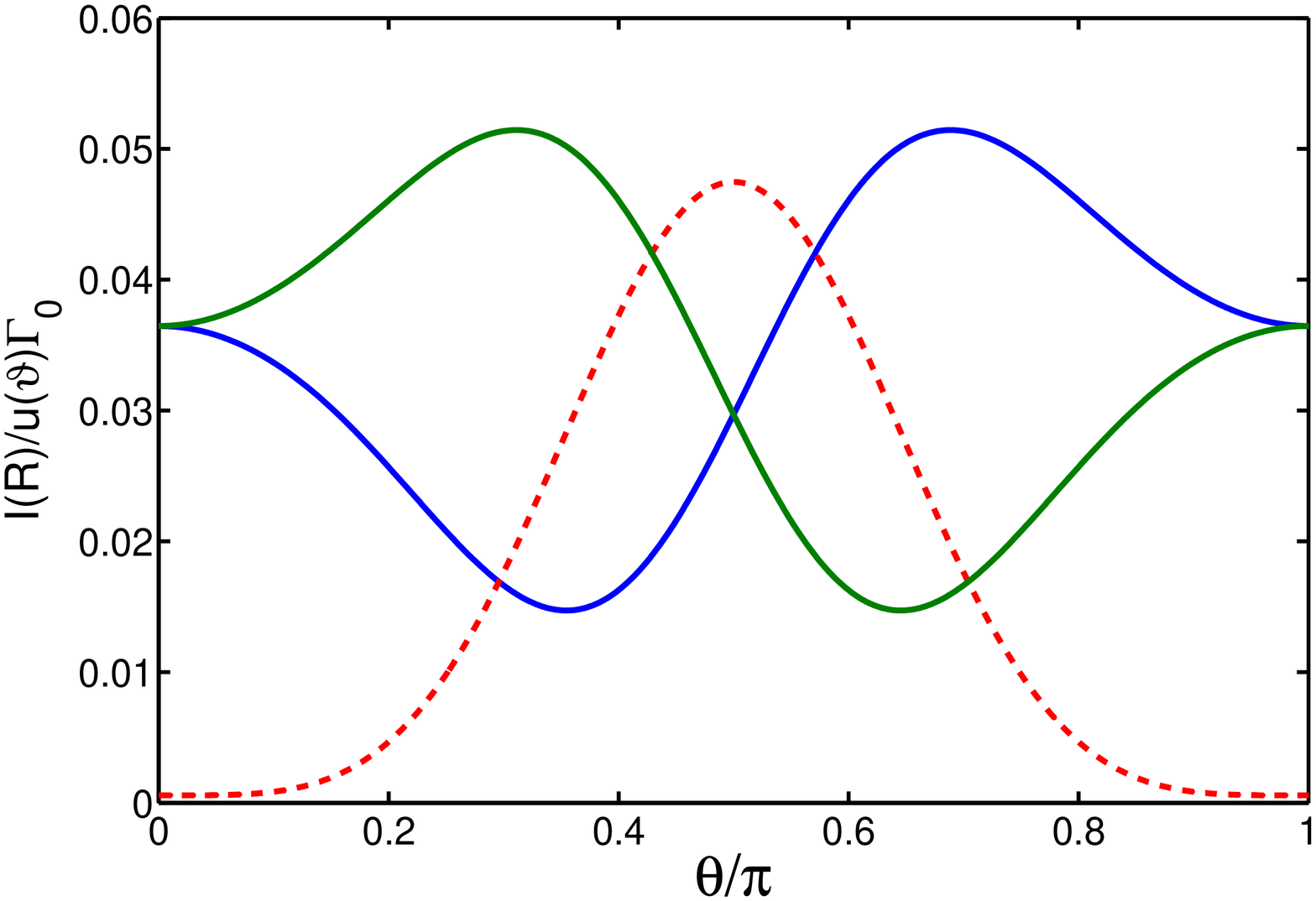}
   \end{tabular}
   \end{center}
   \caption[example]{(Color online) Angular distribution of the radiation intensity for the atomic separation $r_{12} =\lambda/2$, $\Omega=0.2\Gamma_{0}$, $\Delta_{L}=-0.75\Gamma_{0}$, $\Gamma_{1}=\Gamma_{2}$, and different values of $\Delta$: $\Delta =2\Gamma_{0}$ (blue line), $\Delta=-2\Gamma_{0}$ (green line), $\Delta=0$ (red dashed line).}
\label{fig6} 
\end{figure} 

Figure~\ref{fig6} shows the angular distribution of the emitted light intensity for $\gamma=0$ and several values of $\Delta$. According to Eq.~(\ref{e47}), in this case the amplitude of the sine term is entirely governed by $\Delta$. When $\Delta\neq 0$ a peak is seen in a direction deviating significantly from the direction normal to the atomic axis. Depending on the sign of $\Delta$, the peak occurs either in the direction $\theta=\pi/3$ or $\theta=2\pi/3$. These angles correspond to the directions of propagation of the antisymmetric modes (Fig.~\ref{fig2}b). For a positive $\Delta$, so that $\omega_{1}>\omega_{2}$, the intensity of the emitted light is enhanced in the direction $\theta=2\pi/3$ and suppressed in the direction $\theta=\pi/3$. Conversely, for a negative $\Delta$ the maximum of the intensity occurs in the the direction of $\theta=\pi/3$ and a minimum in the direction $\theta=2\pi/3$. Thus, the maximum of the emitted light turns to the right or to the left with respect to the direction normal to the atomic axis when $\Delta\neq 0$. It is interesting that the emitted light turns towards the atom of smaller resonance frequency even if the laser is tuned above the average atomic frequency, $\Delta_{L}=\omega_{0}-\omega_{L} <0$. In fact, the turning direction is independent of the sign of $\Delta_{L}$.

\begin{figure}[h]
   \begin{center}
   \begin{tabular}{c}
   \includegraphics[height=5.5cm]{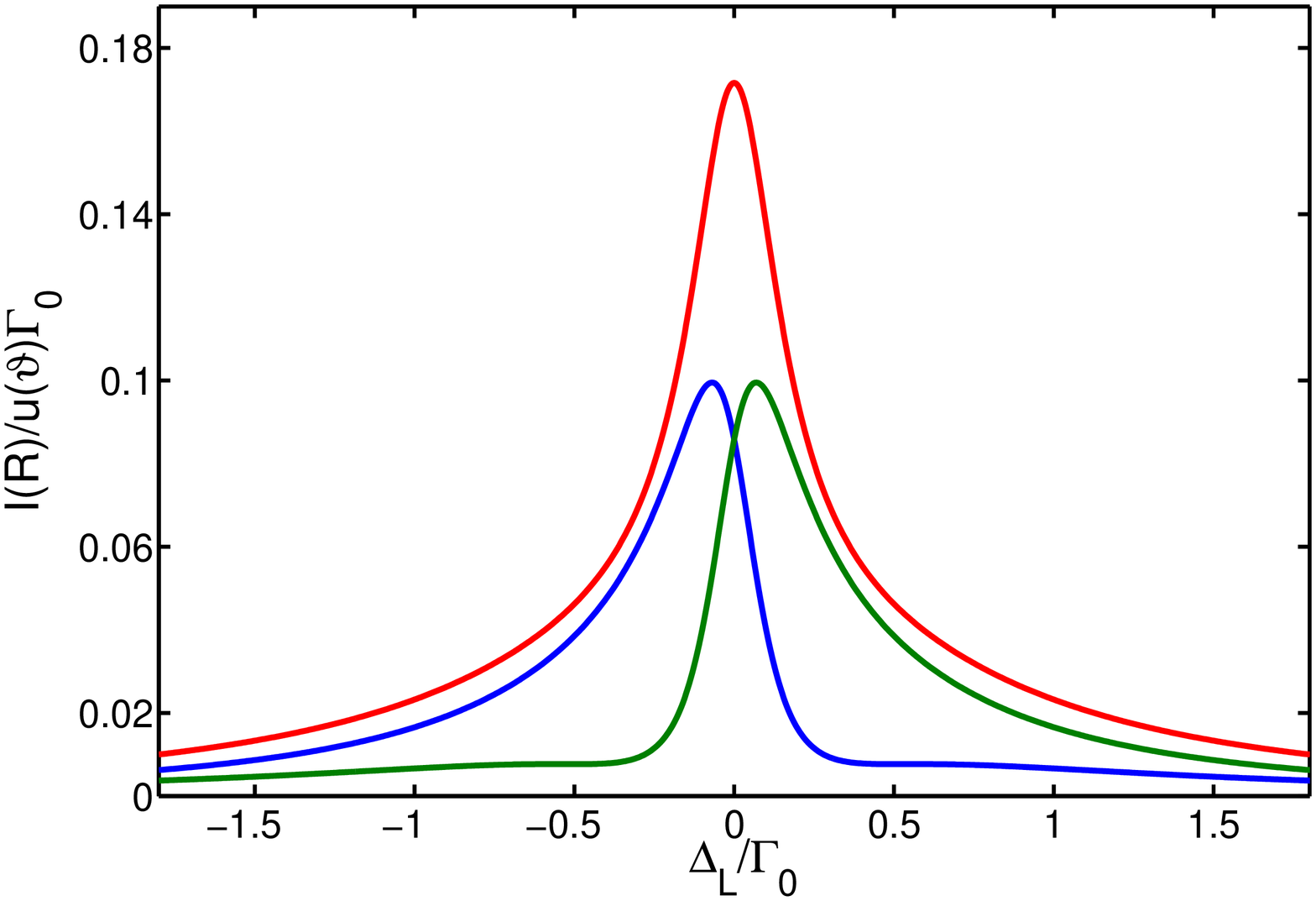}
   \end{tabular}
   \end{center}
   \caption[example]{(Color online) The variation of the steady-state radiation intensity with the laser detuning $\Delta_{L}$ for $\Omega=0.2\Gamma_{0}$, $r_{12}=\lambda/2$, $\Delta=0$, $\Gamma_{2}/\Gamma_{1}=10$ and two different directions of observation, $\theta =\pi/3$ (blue line) and $\theta=2\pi/3$ (green line). The red line is the sum of the two.}
\label{fig7} 
\end{figure}

Figure~\ref{fig7} shows the variation of the steady-state radiation intensity with the laser detuning $\Delta_{L}$ observed in two different directions for atoms with equal resonance frequencies, $\Delta=0$, but different damping rates, $\Gamma_{2}=10\Gamma_{1}$. We have chosen the angles $\theta=\pi/3$ and $\theta=2\pi/3$ which correspond to the directions of propagation of the antisymmetric modes when $r_{12}=\lambda/2$. Note that the intensity of light emitted in the direction $\theta=2\pi/3$ is a mirror image of the intensity emitted in the direction $\theta=\pi/3$. It is apparent that for a given direction of propagation, the intensity is strongly asymmetric about $\Delta_{L}=0$. Namely, for the direction $\theta=\pi/3$, the intensity is large for negative detunings, $\Delta_{L}<0$, but is almost zero for positive detunings, $\Delta_{L}>0$. The situation is completely opposite for the direction $\theta = 2\pi/3$. Now the intensity is large for positive detunings and is almost zero for negative detunings. Thus, by varying the laser frequency from blue detuned $(\omega_{L}>\omega_{0})$ to red detuned $(\omega_{L}<\omega_{0})$ from the atomic resonance, one can switch the emission direction from the mode propagating in the direction $\theta=\pi/3$ to the mode propagating in the direction $\theta=2\pi/3$. Hence, the simple formula in Eq.~(\ref{e47}) predicts clearly that the system of two non-identical two-level atoms may work as a nano-antenna for mode switching and directional light routing.

To distinguish between directions of the emission one can examine the intensity or fringe contrast factor, which for the antisymmetric modes is given by
\begin{align}
 C &= \frac{i\sqrt{1-\gamma^{2}}\!\left(\langle S_{1}^{+}(t)S_{2}^{-}(t)\rangle\!-\!\langle S_{2}^{+}(t)S_{1}^{-}(t)\rangle\right)}{(1+\gamma)\langle S_{1}^{+}(t)S_{1}^{-}(t)\rangle\!+\!(1-\gamma)\langle S_{2}^{+}(t)S_{2}^{-}(t)\rangle} \nonumber\\
 & = \frac{I_{s}(t)}{I_{0}(t)} ,\label{e49}
\end{align}
where $I_{s}(t)$ is the amplitude of the sine term and $I_{0}(t)$ is the amplitude of the background field of the intensity formula~(\ref{e34}). 
 The absolute value $|C|$, called the visibility, determines the relative amplitude between the maxima and minima of the intensity pattern. However, we consider $C$ instead of $|C|$ for a simple reason that the sign of $C$ contains the information about the direction of emission with respect to the direction normal to the atomic axis. For positive values of~$C$, maxima of the intensity occur in directions corresponding to $\sin(kr_{12}\cos\theta)=1$, and for negative $C$ there are minima at these directions. 
   \begin{figure}[h]
   \begin{center}
   \begin{tabular}{c}
   \includegraphics[height=5.5cm]{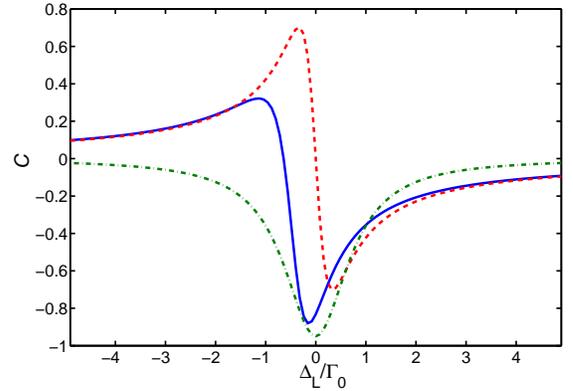}
   \end{tabular}
   \end{center}
   \caption[example]{(Color online) The steady-state intensity contrast factor $C$ as a function of the laser detuning $\Delta_{L}$ for $\Omega=0.2\Gamma_{0}$ and $r_{12}=\lambda/2$. The solid blue line is for $\Delta=\Gamma_{0}, \Gamma_{2}/\Gamma_{1}=10$; the dashed red line is for $\Delta=0, \Gamma_{2}/\Gamma_{1}=10$; and the dashed-dotted green line is for $\Delta=\Gamma_{0}, \Gamma_{2}/\Gamma_{1}=1$.}
\label{fig8} 
\end{figure} 

The contrast factor $C$ is plotted against the laser detuning~$\Delta_{L}$ in Fig.~\ref{fig8} for various detunings $\Delta$ and ratios $\Gamma_{2}/\Gamma_{1}$. For atoms with equal resonance frequencies $(\Delta=0)$ but different damping rates $(\Gamma_{2}>\Gamma_{1})$, the factor $C$ is positive for negative $\Delta_{L}$ and negative for positive $\Delta_{L}$. The threshold at which $C$ changes sign is at $\Delta_{L}=0$. When $\Delta\neq 0$ and $\Gamma_{2}/\Gamma_{1}\neq 1$, the factor is strongly asymmetric and reaches a large negative value, $C\approx -0.9$ at a negative $\Delta_{L}$. The threshold at which $C$ changes signs shifts towards a negative $\Delta_{L}$. For $\Gamma_{2}=\Gamma_{1}$, the factor $C$ is negative for all detunings $\Delta_{L}$. In terms of the directionality of the emission, for $\Delta=0$ and $(\Gamma_{2}>\Gamma_{1})$, the system radiates strongly to the mode $\theta=\pi/3$ for all negative $\Delta_{L}$ with the maximum visibility $|C|=0.7$. The situation reverses for positive $\Delta_{L}$ at which the system radiates strongly to the mode $\theta=2\pi/3$ with the same maximal visibility $0.7$. When $\Delta\neq 0$ and $\Gamma_{2}/\Gamma_{1}\neq 1$, the factor $C$ is strongly asymmetric with large negative values approaching~$-1$ at a small negative $\Delta_{L}$. At that detuning the system radiates only to the mode $\theta=2\pi/3$. For atoms with equal damping rates $(\Gamma_{2}=\Gamma_{1})$ but unequal resonance frequencies $(\Delta\neq 0)$, the factor $C$ is negative for all $\Delta_{L}$ indicating that the directionality of emission cannot be changed by varying the frequency $\omega_{L}$ of the incident laser. 
   \begin{figure}[h]
   \begin{center}
   \begin{tabular}{c}
   \includegraphics[height=5.5cm]{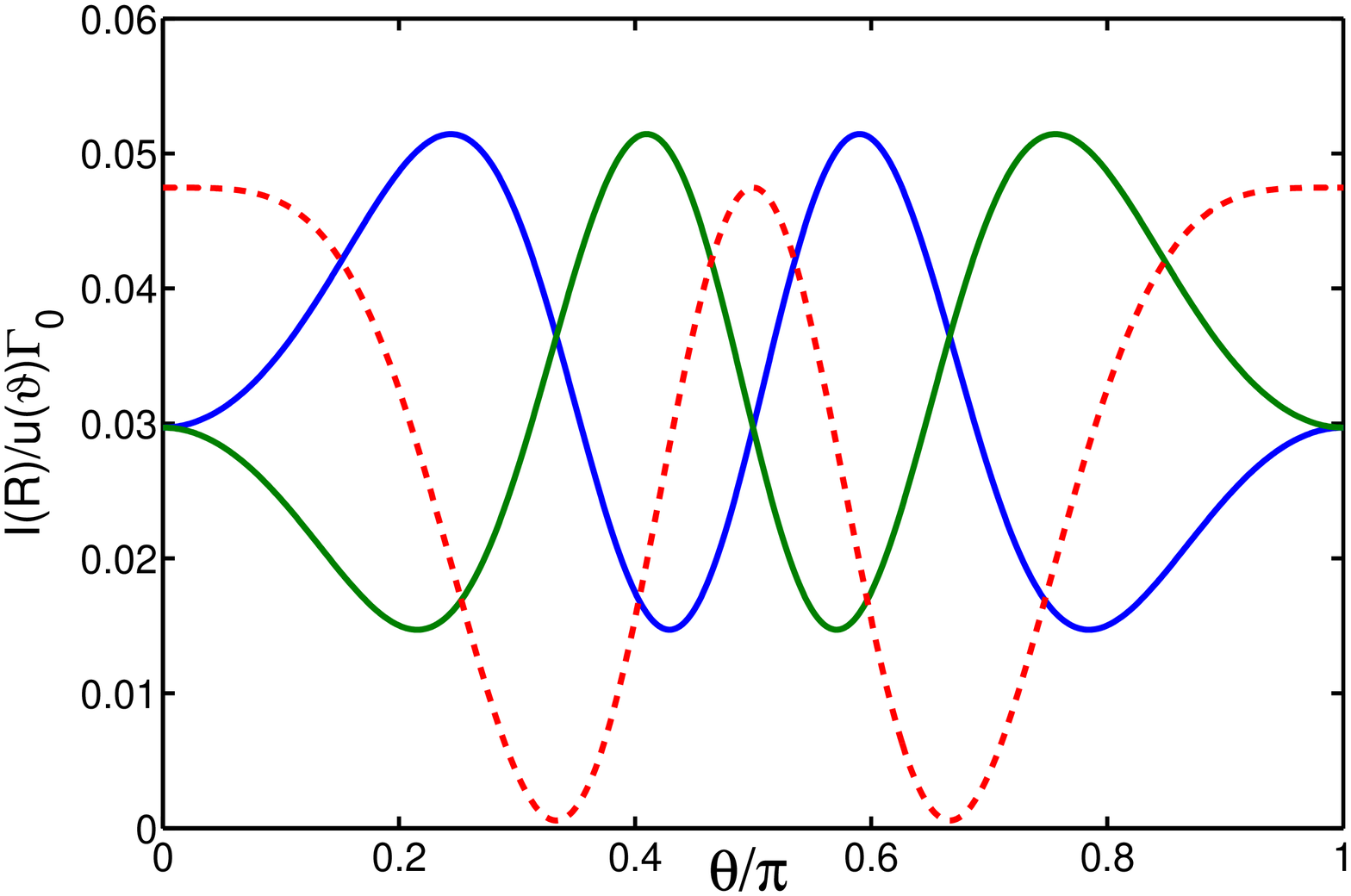}
   \end{tabular}
   \end{center}
   \caption[example]{(Color online) Angular distribution of the emitted field intensity for the atomic separation $r_{12} =\lambda$, $\Omega=0.2\Gamma_{0}$, $\Gamma_{1}=\Gamma_{2}$, $\Delta_{L}=-0.75\Gamma_{0}$ and different values of $\Delta$: $\Delta =2\Gamma_{0}$ (blue line), $\Delta=-2\Gamma_{0}$ (green line), $\Delta=0$ (red dashed line).}
\label{fig9} 
\end{figure} 

In closing this section we point out that the features of mode switching and light routing are similar when distances between the atoms are larger. In Fig.~\ref{fig9} we illustrate the effect of the detuning $\Delta$ on the angular distribution of the emitted field intensity by taking the distance between the atoms $r_{12}=\lambda$, again for the case of equal damping rates $(\Gamma_{1}=\Gamma_{2})$. The effect of $\Delta$ is to switch the emission from the symmetric modes propagating at angles $\theta=0, \pi/2, \pi$ into two antisymmetric modes propagating either at angles $\theta=0.41\pi$ and $0.77\pi$ or $\theta=0.23\pi$ and $0.58\pi$. We can distinguish two characteristic pairs of modes and the emission can be switched from one pair to the other by changing the sign of the detuning $\Delta$. It is interesting, and perhaps surprising, that the system does not turn all of the emitted light into one direction, it rather spits the emitted light into two opposite directions in respect to the direction normal to the atomic axis. In each pair of the modes, one of the modes propagates in a direction~$\theta<\pi/2$ and the other propagates in a direction~$\theta>\pi/2$.

\section{Light routing by interacting atoms}\label{sec8}

We now proceed to illustrate the features of mode switching and light routing fully incorporating the effects of the interaction between the atoms. In order to study these features we numerically evaluate the steady-state values of the density matrix elements, Eq.~(\ref{e43}), that we then apply to graphically display the results for the angular distribution of the radiation intensity and its dependence on the detuning $\Delta_{L}$. In order to work out the effects of the interatomic interactions most clearly we maintain the parameters the same as above for the independent~atoms. 
   \begin{figure}[h]
   \begin{center}
   \begin{tabular}{c}
   \includegraphics[height=5cm]{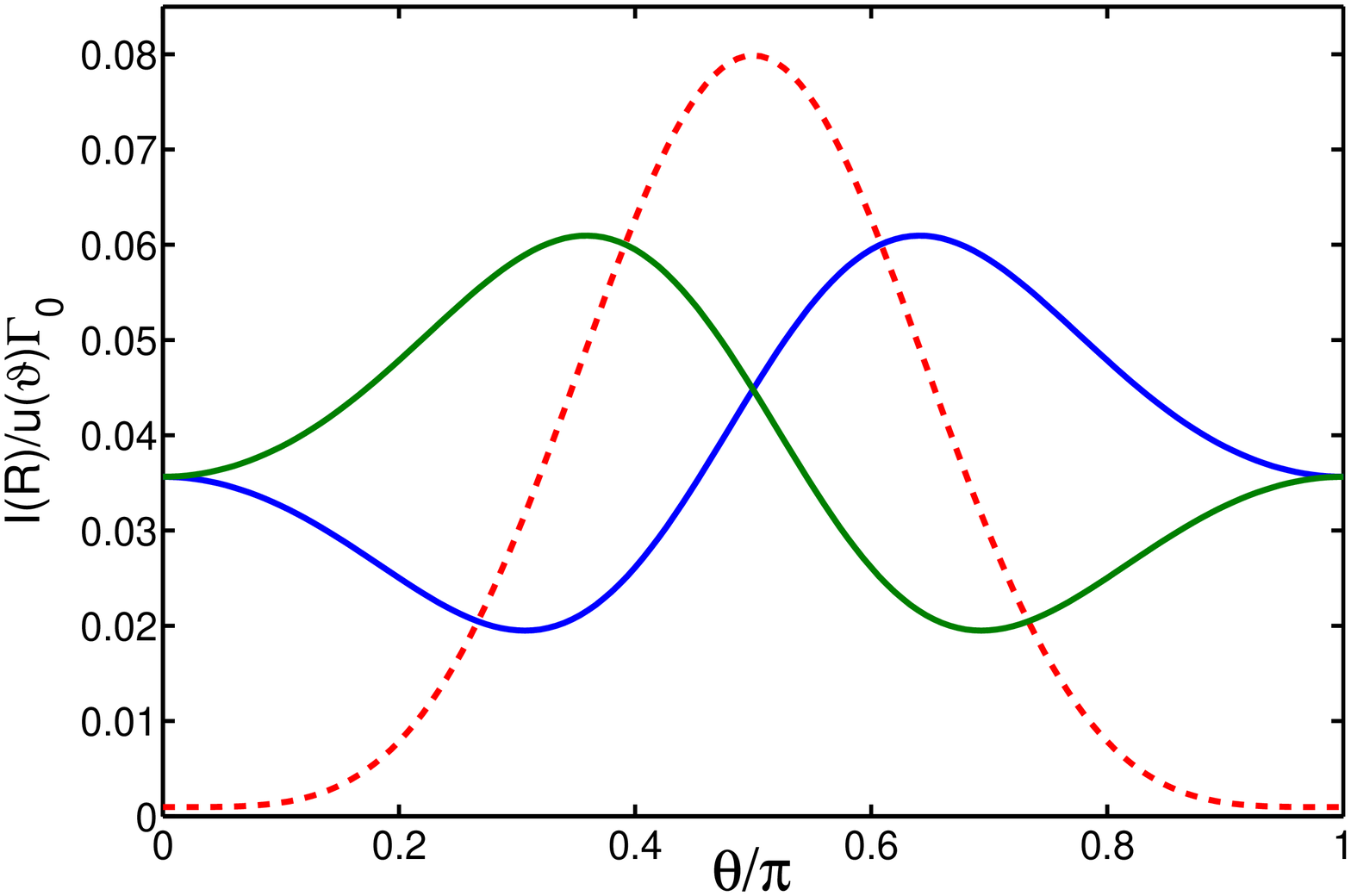}
   \end{tabular}
   \end{center}
   \caption[example]{(Color online) Angular distribution of the radiation intensity for the atomic separation $r_{12} =\lambda/2$ at which $\Gamma_{12}=-0.152\Gamma_{0}$, $\Omega_{12}=0.215\Gamma_{0}$, for $\Omega=0.2\Gamma_{0}$, $\Gamma_{1}=\Gamma_{2}$, $\Delta_{L}=-0.75\Gamma_{0}$ and different values of $\Delta$: $\Delta =2\Gamma_{0}$ (red line), $\Delta=-2\Gamma_{0}$ (blue line), and $\Delta=0$ (dashed red line).}
\label{fig10} 
\end{figure} 

Let us first consider the effect of the interatomic interactions on the switching and light routing for the case of $\Delta\neq 0$ and $\Gamma_{1}=\Gamma_{2}$. Figure~\ref{fig10} shows the angular distribution of the radiation intensity for the interatomic separation $r_{12}=\lambda/2$, equal damping rates, $\Gamma_{1}=\Gamma_{2}$, and different detunings $\Delta$. The parameters are the same as those of Fig.~\ref{fig6}. For identical atoms, $\Delta=0$, the intensity exhibits a peak in the direction normal to the atomic axis, $\theta=\pi/2$, the direction of the symmetric mode. For a positive detuning, the maximum of the intensity is shifted to the direction $\theta=\pi/3$ whereas for a negative $\Delta$ the maximum is shifted to $\theta=2\pi/3$, the directions of the antisymmetric modes. Moreover, the intensity detected at $\theta=\pi/3$ is a mirror image of the intensity detected at $\theta=2\pi/3$. 
   \begin{figure}[h]
   \includegraphics[height=5.5cm]{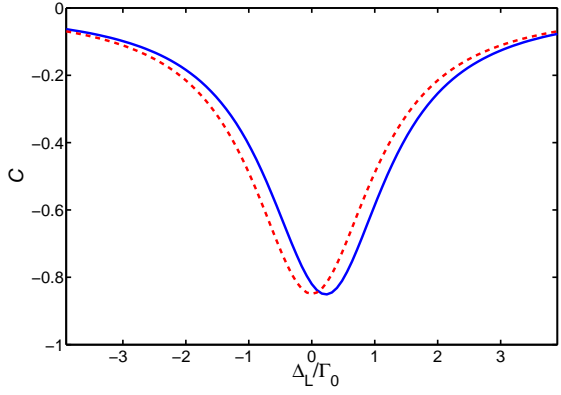}
    \caption[example]{(Color online) The steady-state intensity contrast factor $C$ as a function of the laser detuning $\Delta_{L}$ for $\Omega=0.2\Gamma_{0}$, $r_{12}=\lambda/2$, $\Gamma_{2}=\Gamma_{1}$ and $\Delta=2\Gamma_{0}$. The dashed red line is for independent atoms $(\Omega_{12}=\Gamma_{12}=0)$, and the solid blue line is for interacting atoms with numerical values of the collective parameters $\Gamma_{12}=-0.152\Gamma_{0}$ and $\Omega_{12}=0.215\Gamma_{0}$, evaluated for $r_{12}=\lambda/2$ and~$\hat{\mu}_{1(2)}\perp\vec{r}_{12}$.}
   \label{fig11} 
   \end{figure} 

On comparing Fig.~\ref{fig10} with Fig.~\ref{fig6} no significant differences are present. When the interatomic interactions are included, in Fig.~\ref{fig10}, the angular distribution of the radiation intensity is seen to be qualitatively similar to that in Fig.~\ref{fig6} for independent atoms. More precisely, the interactions slightly alter the visibility by shifting its maximum value to a finite detuning~$\Delta_{L}$. This is shown in Fig.~\ref{fig11}, where we compare the visibility for independent atoms with that for interacting atoms. It is seen that the effect of the interactions is to shift the maximum value of the visibility to a finite $\Delta_{L}$. The magnitude of the shift is equal to the magnitude of the dipole-dipole interaction $\Omega_{12}=0.215\Gamma_{0}$. The shift results in an enhanced visibility for positive $\Delta_{L}$ and a reduced visibility for negative $\Delta_{L}$. Thus, in the case of $\Gamma_{1}=\Gamma_{2}$, the directionality of the emission is not influenced significantly by the interatomic interactions.
   \begin{figure}[h]
   \begin{center}
   \begin{tabular}{c}
   \includegraphics[height=5.5cm]{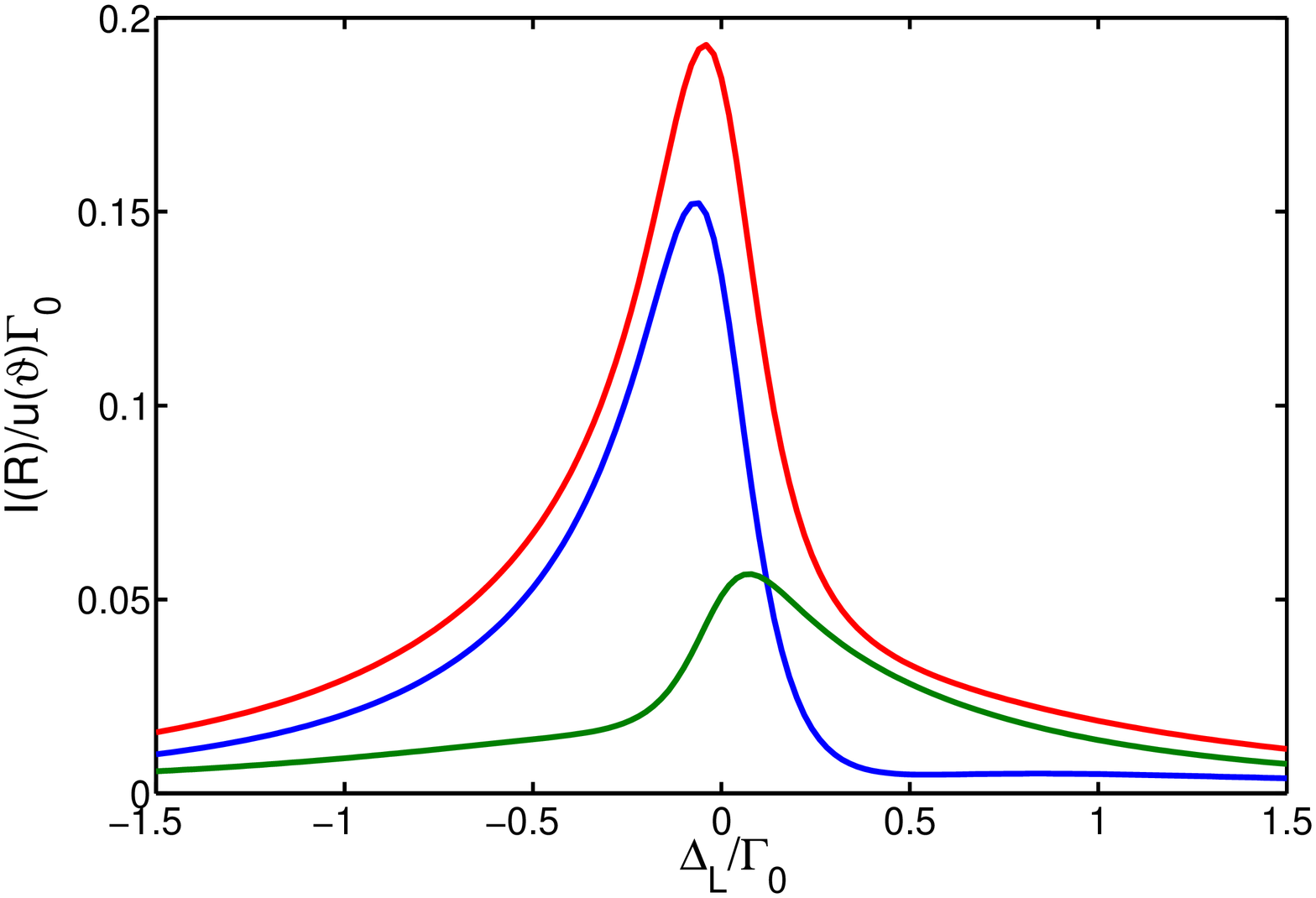}
   \end{tabular}
   \end{center}
   \caption[example]{(Color online) The variation of the steady-state radiation intensity with the laser detuning $\Delta_{L}$ in the presence of the direct interaction between the atoms for $r_{12}=\lambda/2$, $\Omega=0.2\Gamma_{0}$, $\Delta=0$, $\Gamma_{2}/\Gamma_{1}=10$ and the propagation directions of the two antisymmetric modes, $\theta =\pi/3$ (blue line) and $\theta=2\pi/3$ (green line). The red line is the sum of the two.}
\label{fig12} 
\end{figure} 

Let us now turn to the case of $\Gamma_{1}\neq\Gamma_{2}$ and $\Delta=0$, and consider the variation of the radiation intensity with the detuning $\Delta_{L}$. In Fig.~\ref{fig12} the radiation intensity detected in two directions, $\theta=\pi/3$ and $\theta=2\pi/3$, is plotted against $\Delta_{L}$ for the same parameters as in Fig.~\ref{fig7}, but now fully incorporating the interactions between the atoms. The interactions affect the light routing between the antisymmetric modes more drastically than the angular distribution, that the routing is significantly different compared to the case of independent atoms (Fig.~\ref{fig7}). An important difference is that in the present case, the intensity of the mode $\theta=2\pi/3$ is not a mirror image of the intensity of the mode $\theta=\pi/3$. It turns out that the interatomic interactions enhance the emission into the mode propagating in the direction $\theta=\pi/3$ and reduces the emission into the mode propagating in the direction $\theta=2\pi/3$. Since $\Gamma_{2}>\Gamma_{1}$, we may conclude that the dipole-dipole interaction has the effect of turning the emission towards the atom of smaller damping rate. The situation is analogous when the positions of the atoms are interchanged. In this case, the emission is enhanced into the mode propagating in the direction $\theta=2\pi/3$ and significantly reduced in the direction $\theta=\pi/3$.
   \begin{figure}[h]
   \begin{center}
   \begin{tabular}{c}
   \includegraphics[height=5.5cm]{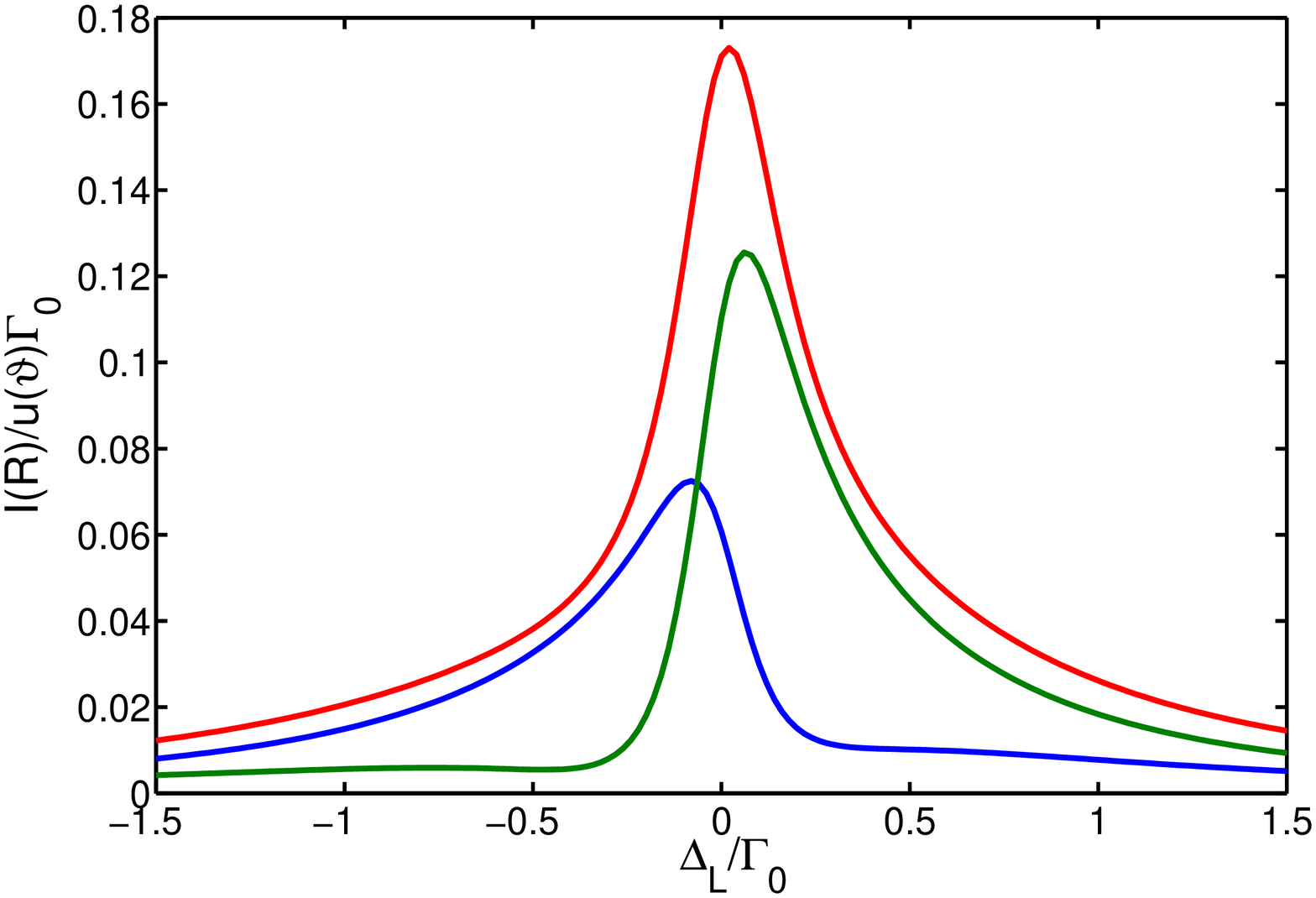}
   \end{tabular}
   \end{center}
   \caption[example]{(Color online) The variation of the steady-state radiation intensity with the laser detuning $\Delta_{L}$ in the presence of the direct interaction between the atoms for $r_{12}=\lambda$, $\Omega=0.2\Gamma_{0}$, $\Delta=0$, $\Gamma_{2}/\Gamma_{1}=10$ and the two pairs of modes propagating at angles $(0.41\pi, 0.77\pi)$ (blue line) and $(0.23\pi, 0.58\pi)$ (green line). The red line is the sum of the two.}
\label{fig13} 
\end{figure} 

The results presented in Fig.~\ref{fig12} could suggest that it is a general feature of the dipole-dipole interaction that in the case of $\Gamma_{2}>\Gamma_{1}$, the interaction enhances emission into modes propagating in a direction $\theta<\pi/2$. This is true for $r_{12}=\lambda/2$, but the situation differs for $r_{12}=\lambda$. Figure~\ref{fig13} shows the corresponding behaviour of the steady-state radiation intensity for $r_{12}=\lambda$. We have seen that the emission can be switched from the symmetric modes to antisymmetric modes where it groups into two pairs of directions, one pair corresponding to $\sin(\pi\cos\theta)=1$ and the other to $\sin(\pi\cos\theta)=-1$.
We see from the figure that in the presence of the dipole-dipole interaction, intensities of the modes corresponding to $\sin(\pi\cos\theta)=1$ are no longer a mirror image of intensities of the modes corresponding to $\sin(\pi\cos\theta)=-1$. The dipole-dipole interaction enhances the emission into the modes corresponding to $\sin(\pi\cos\theta)=-1$ and reduces the emission into the modes corresponding to $\sin(\pi\cos\theta)=1$.
Since the two modes propagating in directions $\theta<\pi/2$ belong to different pairs, only one of the modes is enhanced by the dipole-dipole interaction. The emission into the other mode is reduced by the dipole-dipole interaction. The same conclusion applies to the two modes propagating in directions $\theta>\pi/2$. A careful analysis shows that the dipole-dipole interaction enhances emission into modes whose average propagation angle $\theta_{{\rm av}}<\pi/2$ and reduces emission into modes whose average propagation angle $\theta_{{\rm av}}>\pi/2$. It is easy to see, the emission is enhanced into two modes corresponding to $\sin(\pi\cos\theta)=-1$ and reduced in modes corresponding to $\sin(\pi\cos\theta)=1$.

We may conclude that the dipole-dipole interaction has a significant impact on mode switching and light routing when~$\Gamma_{2}\neq\Gamma_{1}$. It has the effect to turn the emission towards the atom of smaller damping rate. However, the general conclusion of the case of independent atoms remains unchanged that a blue detuned laser field will direct the emitted light to modes corresponding to $\sin(kr_{12}\cos\theta)=1$, but a red detuned field will direct the emission to modes corresponding to~$\sin(kr_{12}\cos\theta)=-1$.
   \begin{figure}[h]
   \begin{center}
   \begin{tabular}{c}
   \includegraphics[height=5.5cm]{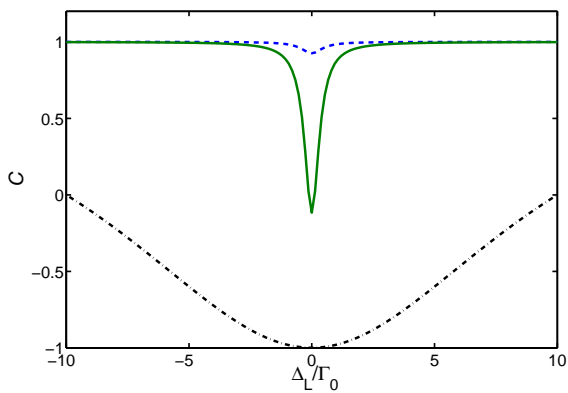}
   \end{tabular}
   \end{center}
   \caption[example]{(Color online) The contrast factor $C$ of the symmetric modes, Eq.~(\ref{e50}), plotted as a function of the laser detuning $\Delta_{L}$ for $\Omega=0.2\Gamma_{0}$, $r_{12}=\lambda/4$, $\Gamma_{2}=\Gamma_{1}$, and different $\Delta$: $\Delta=0$ (dashed blue line), $\Delta=0.5\Gamma_{0}$ (solid green line), $\Delta=20\Gamma_{0}$ (dashed-dotted black line).}
\label{fig14} 
\end{figure} 

We close this section with a brief comment about the physical meaning of negative values of the contrast factor $C$. In a paper by Mayer and Yeoman~\cite{my97} the contrast factor for the symmetric modes was evaluated, which in our notation is
\begin{align}
 C = \frac{\sqrt{1-\gamma^{2}}\left(\langle S_{1}^{+}(t)S_{2}^{-}(t)\rangle + \langle S_{2}^{+}(t)S_{1}^{-}(t)\rangle\right)}{(1+\gamma)\langle S_{1}^{+}(t)S_{1}^{-}(t)\rangle + (1-\gamma)\langle S_{2}^{+}(t)S_{2}^{-}(t)\rangle} .\label{e50}
\end{align}
The authors have considered a system composed of two identical atoms $(\Delta=\gamma=0)$ driven by an incoherent field and simultaneously coupled to a cavity mode, and have found that the factor $C$ can have negative values. The negative values of $C$ indicate a minimum of the radiation intensity to occur in the direction normal to the atomic axis. The fact that~$C$ can have negative values was interpreted as "intrinsically quantum-mechanical effect with no classical analog". We have shown that also the contrast factor of the antisymmetric modes, Eq.~(\ref{e49}), can have negative values, see Figs.~\ref{fig8} and~\ref{fig11}. Similarly, it is not difficult to show that for the system considered in the present paper, also the factor (\ref{e50}) can have negative values. For example, Fig.~\ref{fig14} shows the contrast factor (\ref{e50}) for the situation presented in Fig.~\ref{fig4} of switching the radiation between the symmetric and antisymmetric modes for $r_{12}=\lambda/4$ and $\Delta\neq 0$. It is seen that in the case of nonidentical atoms, the factor (\ref{e50}) can have negative values and at large detunings~$\Delta$ it reaches the optimum negative value $C=-1$. As we have seen in Fig.~\ref{fig4}, at large~$\Delta$ the excitation is completely transferred from the the symmetric to the antisymmetric modes. Thus, our results show that~$C<0$ is not an intrinsically nonclassical effect. It can be interpreted as resulting from a complete transfer of the excitation from the symmetric to the antisymmetric modes that radiate in directions different than normal to the atomic axis.

\section{Conclusions}\label{sec9}

We have investigated radiative properties of a system composed of two nonidentical two-level atoms, especially to show that the system could work as a nano-antenna for the mode switching and light routing. We have analysed different contributions to the radiative intensity from the collective states populations and coherences and have found the coherence between the symmetric and antisymmetric states is crucial for the modes switching and light routing. It has been shown that as long as the atoms are identical, the emission cannot be switched between the symmetric and antisymmetric modes. The switching may occur when the atoms are nonidentical with either different resonance frequencies or different damping rates. In this case, the emission can be routed to different modes by changing the relative ratio of the resonance frequencies, or the ratio of the damping rates or by a proper tuning of the laser frequency to the atomic resonance frequencies. In the case of atoms of different resonance frequencies but equal damping rates, the light routing is independent of the frequency of the driving laser field. It depends only on the sign of the detuning between the atomic resonance frequencies. In contrast, if the atoms have different damping rates, the emission direction can be switched between different modes by changing the laser frequency from the blue to red detuned from the atomic resonance.

We have also considered the effect of the interatomic interactions, in particular, the dipole-dipole interaction on the feature of light routing. While the light routing by the system of interacting atoms with different resonance frequencies is quite similar to that of independent atoms, the system of interacting atoms with different damping rates exhibits an interesting feature that the light routing becoming asymmetric under the dipole-dipole interaction with the enhanced emission into modes turned towards the atom of smaller damping~rate.

Finally, we would like to point out that most of the results obtained in the present paper are closely related to the results of a recent experiment by Shegai {\it et al.}~\cite{she11}, where light routing by a bimetallic nano-antenna consisting of two metallic particles of different plasmon resonances was observed.

\end{document}